\documentclass[journal,twocolumn]{IEEEtran}
\usepackage{diagbox} 
\usepackage{graphicx}
\usepackage{epstopdf}
\usepackage{psfrag}
\usepackage{subfigure}
\usepackage{url}
\usepackage{stfloats}
\usepackage{amsfonts,amssymb,amsmath,bm,paralist,theorem,cite,ifthen,color,nccmath}
\usepackage{caption}
\usepackage{calc}
\usepackage{enumerate}
\usepackage{multirow}
\usepackage{makecell}
\usepackage[ruled]{algorithm2e}
\usepackage{setspace}
\usepackage{array}
\usepackage{float}
\usepackage{soul}

\usepackage[colorlinks,
            linkcolor=red,
            anchorcolor=red,
            citecolor=red]{hyperref}
\usepackage[]{mdframed}

\usepackage[figurename=Fig., labelsep=period, font=small]{caption}

\SetKwInOut{Input}{Input}
\SetKwInOut{Output}{Output}

\newcommand{\T}{{\scriptscriptstyle\mathsf{T}}}
\renewcommand{\H}{{\scriptscriptstyle\mathsf{H}}}

\usepackage[normalem]{ulem}

\newtheorem{definition}{Definition}

\newtheorem{proposition}{Proposition}
\newtheorem{remark}{Remark}
\newtheorem{lemma}{Lemma}
\newtheorem{corollary}{Corollary}

\newcommand{\qedsymbol}{\hfill \(\blacksquare\)}
\graphicspath{{Figures/}}

\newcounter{algoline}

\AtBeginEnvironment{algorithm}{\setcounter{algoline}{0}}
\newcommand\Ccl{\ensuremath{\mathcal{C}}}

\newcommand\Ncl{\ensuremath{\mathcal{N}}}
\newcommand\Ocl{\ensuremath{\mathcal{O}}}

\newcommand\Tcl{\ensuremath{\mathcal{T}}}
\newcommand\Fcl{\ensuremath{\mathcal{F}}}

\newcommand\Cs{\ensuremath{{\mathbb{C}}}}
\newcommand\Es{\ensuremath{{\mathbb{E}}}}

\newcommand\Ab{\ensuremath{ \mathbf{A} }}

\newcommand\Cb{\ensuremath{ \mathbf{C} }}

\newcommand\Eb{\ensuremath{ \mathbf{E} }}
\newcommand\Fb{\ensuremath{ \mathbf{F} }}
\newcommand\Gb{\ensuremath{ \mathbf{G} }}
\newcommand\Hb{\ensuremath{ \mathbf{H} }}
\newcommand\Ib{\ensuremath{ \mathbf{I} }}

\newcommand\Pb{\ensuremath{ \mathbf{P} }}
\newcommand\Ub{\ensuremath{ \mathbf{U} }}
\newcommand\Qb{\ensuremath{ \mathbf{Q} }}
\newcommand\Vb{\ensuremath{ \mathbf{V} }}
\newcommand\Wb{\ensuremath{ \mathbf{W} }}
\newcommand\Jb{\ensuremath{ \mathbf{J} }}
\newcommand\Kb{\ensuremath{ \mathbf{K} }}
\newcommand\Lb{\ensuremath{ \mathbf{L} }}

\newcommand\Zb{\ensuremath{ \mathbf{Z} }}

\newcommand\ab{\ensuremath{ \mathbf{a} }}
\newcommand\bb{\ensuremath{ \mathbf{b} }}

\newcommand\eb{\ensuremath{ \mathbf{e} }}
\newcommand\gb{\ensuremath{ \mathbf{g} }}

\newcommand\nb{\ensuremath{ \mathbf{n} }}

\newcommand\qb{\ensuremath{ \mathbf{q} }}
\newcommand\ssb{\ensuremath{ \mathbf{s} }}

\newcommand\xb{\ensuremath{ \mathbf{x} }}
\newcommand\yb{\ensuremath{ \mathbf{y} }}
\newcommand\zb{\ensuremath{ \mathbf{z} }}

\newcommand\Gammab{\ensuremath{{\bm \Gamma}}}

\newcommand\Sigmab{\ensuremath{{\bm \Sigma}}}

\newcommand\etab{\ensuremath{{\bm \eta}}}

\newcommand\bmax{\ensuremath{b_{\rm max}}}

\newcommand\diag{\ensuremath{{\rm diag}}}
\newcommand\tr{\ensuremath{{\rm Tr}}}

\newcommand{\SNR}{\textrm{SNR}}

\newcommand\dB{\textrm{dB}}
\newcommand\mW{\textrm{mW}}

\newcommand\sgn{\ensuremath{{\rm sgn}}}

\newcommand\Nt{\ensuremath{ N_{\rm t} }}
\newcommand\Nr{\ensuremath{ N_{\rm r} }}

\newcommand\Ns{\ensuremath{ N_{\rm s} }}

\newcommand\Nb{\ensuremath{ N_{\rm b} }}

\newcommand\Pt{\ensuremath{ P_{\rm t} }}

\newcommand\Ncls{\ensuremath{ N_{\rm cl} }}
\newcommand\Nray{\ensuremath{ N_{\rm ray} }}

\newcommand\Nq{\ensuremath{ N_{\rm q} }}
\newcommand\cy{\ensuremath{ c^{\rm y} }}
\newcommand\cx{\ensuremath{ c^{\rm x} }}
\newcommand\ty{\ensuremath{ t^{\rm y} }}
\newcommand\tx{\ensuremath{ t^{\rm x} }}
\newcommand\sigmay{\ensuremath{ \sigma_{\rm y} }}

\newcommand\sigman{\ensuremath{ \sigma_{\rm n} }}
\newcommand\Qy{\ensuremath{ Q_{\rm y} }}
\newcommand\Qx{\ensuremath{ Q_{\rm x} }}

\newcommand\nondiag{\ensuremath{{\rm nondiag}}}

\usepackage{relsize}

 \usepackage{afterpage}

\usepackage{etoolbox}
\newcommand{\zerodisplayskips}{%
  \setlength{\abovedisplayskip}{3pt}%
  \setlength{\belowdisplayskip}{3pt}%
  \setlength{\abovedisplayshortskip}{3pt}%
  \setlength{\belowdisplayshortskip}{3pt}}
\appto{\normalsize}{\zerodisplayskips}
\appto{\small}{\zerodisplayskips}
\appto{\footnotesize}{\zerodisplayskips}

\usepackage{pifont}
\newcommand{\xmark}{\ding{55}} 

\IEEEoverridecommandlockouts

\begin{document}

\title{Joint Beamforming Design and Bit Allocation in Massive MIMO with Resolution-Adaptive ADCs}

\author{Mengyuan~Ma,~\IEEEmembership{Student Member,~IEEE}, Nhan~Thanh~Nguyen,~\IEEEmembership{Member,~IEEE},\\ Italo~Atzeni,~\IEEEmembership{Senior~Member,~IEEE}, and~Markku~Juntti,~\IEEEmembership{Fellow,~IEEE}

\thanks{The authors are with the Centre for Wireless Communications, University of Oulu, Finland (e-mail: \{mengyuan.ma, nhan.nguyen, italo.atzeni, markku.juntti\}@oulu.fi). This work was supported by the Research Council of Finland (332362 EERA, 336449 Profi6, 348396 HIGH-6G, and 357504 EETCAMD, and 369116 6G~Flagship).}
\vspace{-3mm}
}

\maketitle

\begin{abstract}
Low-resolution analog-to-digital converters (ADCs) have emerged as a promising technology for reducing power consumption and complexity in massive multiple-input multiple-output (MIMO) systems while maintaining satisfactory spectral and energy efficiencies (SE/EE). In this work, we first present the fundamental properties of optimal quantization and leverage them to derive a more accurate approximation of the covariance matrix of the quantization distortion. This theoretical finding facilitates the analysis of the system's SE in the presence of low-resolution ADCs. Then, considering resolution-adaptive ADCs, we focus on the joint optimization of the transmit-receive beamforming and bit allocation to maximize the SE under constraints on the transmit power and the total number of active ADC bits. To solve the resulting mixed-integer problem, we first develop an efficient beamforming design for fixed ADC resolutions. Subsequently, we propose a low-complexity heuristic algorithm to iteratively optimize the ADC resolutions and beamforming matrices. Numerical results for a $64 \times 64$ MIMO system demonstrate that the proposed design offers $6\%$ improvements in both SE and EE with $40\%$ fewer active ADC bits compared with uniform bit allocation. Furthermore, it is unveiled that receiving more data streams with low-resolution ADCs can lead to higher SE and EE compared with receiving fewer data streams with high-resolution ADCs.
\end{abstract}


%
\IEEEpeerreviewmaketitle

\section{Introduction}

Massive multiple-input multiple-output (MIMO) is a crucial physical-layer technology for wireless communications at both sub-6GHz and millimeter-wave (mmWave) frequencies \cite{heath2016overview}, addressing the increasing demand for high data rates \cite{jiang2021road}. The large number of antenna elements in massive MIMO provides significant spatial multiplexing gains through beamforming techniques. Digital beamforming (DBF) architectures, which deploy a dedicated radio-frequency (RF) chain for each antenna element, can enable high spectral efficiency (SE) but incur substantial energy costs due to power-intensive RF components, especially analog-to-digital converters (ADCs). For instance, a high-speed ADC operating at $1$~Gsample/s with high resolution (e.g., $8$--$12$ bits) can consume several Watts \cite{li2017channel}. Furthermore, its power consumption increases linearly with the signal bandwidth and exponentially with the number of resolution bits \cite{murmann2015race}, posing a significant challenge to the system's energy efficiency (EE). Consequently, the integration of low-resolution ADCs and DBF has emerged as an effective strategy to curtail power consumption without unduly compromising the SE \cite{liu2019low}.

Another attractive solution in this regard is to utilize hybrid beamforming (HBF) architectures, where a small number of RF chains is connected to the antenna array through a network of phase shifters or switches \cite{mendez2016hybrid}. However, HBF architectures have limited multiplexing capabilities and strongly depend on the calibration of the analog components \cite{roth2018comparison}. Consequently, DBF requires lower circuit cost to achieve a SE similar to that of HBF \cite{yan2019performance}, which makes the former more energy efficient, especially when using low-resolution ADCs \cite{roth2018comparison,castaneda2021resolution}. Moreover, while the water-filling (WF) power allocation achieves the capacity of a full-resolution MIMO system with perfect channel state information (CSI) at both the transmitter and receiver \cite{tse2005fundamentals}, it becomes suboptimal in the presence of low-resolution ADCs, necessitating a more efficient design. 

\subsection{Prior Works}

Recent years have witnessed a proliferation of studies on massive MIMO with
low-resolution ADCs (with $2$--$4$ bits). It has been shown in \cite{singh2009limits,jacobsson2017throughput} that a system using very few bits can approach the performance of a full-resolution one. Mezghani {\it et al.} \cite{mezghani2012capacity} derived a closed-form lower bound for the capacity of a point-to-point MIMO system. More recent works have focused on beamforming designs \cite{mezghani2009transmit,jacobsson2017quantized,ling2019performance,ma2024digital}. Furthermore, mixed-ADC systems, which simultaneously deploy one-bit and high-resolution ADCs, have been shown to perform better than fixed-resolution architectures, especially at high SNR \cite{zhang2016mixed,zhang2017performance,pirzadeh2018spectral}. 
On the other hand, resolution-adaptive ADCs have been studied in \cite{bai2013optimization,ahmed2017joint,choi2017resolution,nguyen2020energy,prasad2020optimizing,castaneda2021resolution,castaneda2021spawc,sheng2020energy,verenzuela2021optimal,verenzuela2017per} to flexibly balance the SE-EE tradeoff of low-resolution systems. For instance, it has been shown in \cite{bai2013optimization,ahmed2017joint,choi2017resolution,nguyen2020energy,sheng2020energy,verenzuela2021optimal,verenzuela2017per} that efficient bit allocation strategies can offer a higher EE compared with uniform-resolution architectures. Moreover, Casta{\~n}eda {\it et al.} \cite{castaneda2021resolution} developed a resolution-adaptive fully digital receiver within an application-specific integrated circuit (ASIC). Furthermore, they demonstrated that a $256$-antenna base station with resolution-adaptive ADCs serving $16$ users allows a reduction in power consumption by $6.7$ times compared with a traditional fixed-resolution design \cite{castaneda2021spawc}. Additional SE gains can be achieved by jointly optimizing the transmit power and ADC resolutions \cite{prasad2020optimizing}. 

Among the aforementioned works, two primary methods are used to model quantization, i.e., the additive quantization noise model (AQNM) \cite{gersho2012vector,fletcher2007robust,orhan2015low} and the Bussgang decomposition \cite{bussgang1952crosscorrelation}. Both approaches approximate the (nonlinear) quantization function using a linear model. However, in the literature, there are two distinct linear approximations referred to as the AQNM. The first is \cite{gersho2012vector}
\begin{equation}\label{eq:linear model 1}
  Q(X)=X+q,
\end{equation}
where $Q(\cdot)$ and $q$ denote the quantization function and {\it quantization error}, respectively. The second is \cite{fletcher2007robust}
\begin{equation}\label{eq:linear model 2}
 Q(X)=\alpha X+ \eta,
\end{equation}
where $\alpha$ is a constant depending on the quantizers and on the distribution of $X$, and $\eta$ represents the {\it quantization distortion} (QD). Both \eqref{eq:linear model 1} and \eqref{eq:linear model 2} can be employed to analyze the worst-case system's performance \cite{diggavi2001worst,hassibi2003much} assuming that $q$ or $\eta$ is a Gaussian random variable uncorrelated with $X$. Model \eqref{eq:linear model 2} was first derived in \cite{fletcher2007robust} and named AQNM later in \cite{orhan2015low}; it was also called the pseudo-quantization noise model in \cite{zhang2016mixed}. Although \eqref{eq:linear model 2} and the Bussgang decomposition were developed from separate technical lineages, it was shown in \cite{demir2020bussgang} that the former is the latter tailored for the case of quantization. Therefore, we call the model in \eqref{eq:linear model 2} as the \emph{Bussgang-based AQNM} (BAQNM) while we refer to \eqref{eq:linear model 1} as the AQNM for distinction. The AQNM is typically less accurate than the BAQNM because the assumption that $q$ is uncorrelated with $X$ is generally not satisfied. In contrast, $\eta$ is uncorrelated with $X$ based on the properties of the Bussgang decomposition. Furthermore, the QD covariance matrix is a key ingredient for the performance analysis and optimization with the BAQNM. A diagonal approximation of the QD covariance matrix was derived in \cite{mezghani2012capacity,bai2013optimization}, which has since then been widely used in the literature. Nonetheless, this approximation can introduce substantial error in the performance analysis and optimization in some scenarios \cite{demir2020bussgang, prasad2020optimizing}. It has been numerically demonstrated that, at low SNR, using the BAQNM and the approximation of the QD covariance matrix can obtain a SE close to the channel capacity \cite{orhan2015low,roth2017achievable}. However, at high SNR, the diagonal approximation of the QD covariance matrix results in non-negligible performance overestimation, especially for massive MIMO systems with very few ADC bits \cite{roth2017achievable,demir2020bussgang}. 


\subsection{Contributions}\label{eq:sec contribution}
 Recently, Casta{\~n}eda {\it et. al} \cite{castaneda2021spawc,castaneda2021resolution} implemented the first fully digital systems with resolution-adaptive ADCs, where the resolution bits are dynamically adjusted to adapt to the instantaneous communications scenario, e.g., CSI and modulation scheme. They demonstrated that the adoption of resolution-adaptive ADCs can achieve power savings with several orders of magnitude for realistic mmWave channels, underscoring the significant potential of this technology. Previous research on resolution-adaptive ADCs has mainly focused on bit allocation \cite{bai2013optimization, choi2017resolution, nguyen2020energy,verenzuela2021optimal,verenzuela2017per} or the combined optimization of bit allocation with either transmit or receive beamforming design \cite{ahmed2017joint,prasad2020optimizing, sheng2020energy}. However, the joint optimization of transmit-receive beamforming and bit allocation holds significant potential for achieving higher SE and providing valuable insights into the SE-EE tradeoff, yet it remains largely unexplored. Moreover, existing designs \cite{bai2013optimization, choi2017resolution, nguyen2020energy,ahmed2017joint, prasad2020optimizing, sheng2020energy,verenzuela2021optimal,verenzuela2017per} fail to adequately address the complex coupling between transmit-receive beamformers and the bit allocation vector, which demands a comprehensive and integrated design approach. This work bridges this critical gap by introducing a joint transmit-receive beamforming design and bit allocation in point-to-point MIMO systems employing resolution-adaptive ADCs.
 
 To facilitate the design, it is essential to properly model the quantization process. Although the BAQNM is widely adopted in the literature, its foundational assumptions, which significantly influence its accuracy, have not been thoroughly examined. Existing studies mainly use the BAQNM as a plug-and-play tool without evaluating its applicability, which may lead to questionable conclusions. To address this issue, we aim to develop a more accurate quantization model that enables reliable and insightful analysis. Table~\ref{tb:Comparison of existing research on resolution-adaptive ADCs response} outlines the distinctions between our work
and prior studies.
 
 
 The specific contributions of this paper are summarized as follows.

\begin{table}[tb]
\small
\renewcommand\arraystretch{1.5}
\centering
\caption{Comparison of prior works on resolution-adaptive ADCs.}
\label{tb:Comparison of existing research on resolution-adaptive ADCs response}
    \begin{tabular}{|p{1.4cm}|p{1.3cm}|p{1.7cm}|p{1.7cm}|p{1.7cm}|} 
    \hline   
     Reference  & \parbox{3cm}{Bit \\ allocation} &   \parbox{3cm}{Transmit \\ beamforming} & \parbox{3cm}{Receive \\ beamforming} &  \parbox{3cm}{Quantization \\modeling\\analysis}  \\
      \hline
      \hline       \cite{bai2013optimization,choi2017resolution,nguyen2020energy,verenzuela2021optimal,verenzuela2017per}      & \checkmark  & \xmark & \xmark& \xmark\\
    \hline  
     \cite{prasad2020optimizing} & \checkmark  & \checkmark & \xmark &\xmark\\
    \hline 
    \cite{ahmed2017joint,sheng2020energy}  & \checkmark  & \xmark &  \checkmark &\xmark\\
        \hline 
    This work  & \checkmark  & \checkmark &  \checkmark  &  \checkmark   \\
    \hline 
    \end{tabular}   
\end{table}

\begin{itemize}
\item We first present the fundamental properties of optimal quantization, including the scaling law, distortion invariance, and some important statistical properties between random variable and its quantized output. The scaling law enables efficient derivation of the optimally quantized outputs for Gaussian signals with different variances, significantly reducing computational and time complexities. As a result, it allows to efficiently obtain the actual QD covariance matrix via numerical methods, which includes the non-zero off-diagonal entries. This leads to more reliable performance evaluation compared to that solely relying on the theoretical diagonal approximation. In addition, the distortion invariance property ensures that the Bussgang gain matrix in the BAQNM depends solely on the distortion factor, which is determined by the quantizer resolution. We propose a more accurate expression of this distortion factor, enabling reliable performance evaluation when employing the BAQNM. 

\item Leveraging the fundamental properties of optimal quantization and the Bussgang decomposition, we obtain a more accurate characterization of the BAQNM and the diagonal approximation of the QD covariance matrix compared to those developed in \cite{mezghani2012capacity,bai2013optimization}. Our analysis shows that the BAQNM and the diagonal approximation of the QD covariance matrix only hold when Gaussian signals are optimally quantized. Moreover, we reveal the connections and nuances between applying BAQNM and the arcsine law \cite{jacovitti1994estimation} to one-bit quantization. The BAQNM applies to only the optimal quantization, whereas the arcsine law implies more general quantization for one-bit systems.

\item Building upon the theoretical findings explained above, we consider the joint transmit-receive beamforming design and bit allocation problem to maximize the SE subject to constraints on the transmit power budget and the total number of active ADC bits. This design problem is inherently complex due to its mixed-integer nature. We address it by first determining the beamformer under fixed ADC resolutions. Subsequently, we propose a low-complexity algorithm to iteratively optimize the ADC resolutions and the beamforming matrices.

\item We perform extensive numerical simulations based on the simulated QD covariance matrix, which has non-zero off-diagonal entries, using the scaling law of optimal quantization. The results confirm the superiority of the proposed schemes over the state-of-the-art algorithms, especially beamforming alone with $2$--$4$ ADC bits per RF chain. For example, in a $64 \times 64$ MIMO system, the proposed design offers $6\%$ improvements in both SE and EE while requiring $40\%$ fewer active ADC bits compared with uniform bit allocation. Moreover, the SE-EE comparison shows that receiving more data streams with low-resolution ADCs can achieve higher SE and EE than receiving fewer data streams with high-resolution ADCs. 
\end{itemize}


\subsection{Organization and Notations}

The rest of this paper is organized as follows. In Section~\ref{sec:system model}, we present the system and quantization models. The BAQNM and the approximation of the QD covariance matrix are then derived in Section~\ref{sec:QD approx}. We delve into the joint transmit-receive beamforming and bit allocation design in Section~\ref{sec:transceiver design}. Finally, we provide simulation results and conclusions in Sections~\ref{sec:simulation} and~\ref{sec:conclusion}, respectively.

Scalars, vectors, and matrices are denoted by the lower\-case, boldface lowercase, and boldface uppercase letters, respectively. Furthermore, we use $\left(\cdot\right)^*$, $\left(\cdot\right)^\T$, $\left(\cdot\right)^\H$, and $\left(\cdot\right)^{-1}$ to represent the conjugate, transpose, conjugate transpose, and matrix inverse operators, respectively.  $\left\|\cdot\right\|_{\Fcl}$ signifies the Frobenius norm for matrices, whereas $\otimes$ denotes the Kronecker product. In addition, the expectation and trace operators are represented by $\Es\left(\cdot\right)$ and $\tr\left(\cdot\right)$. We use $\left|a \right|$ and $\det \left(\Ab \right)$ to denote the absolute value of the scalar $a$ and the determinant of the matrix $\Ab$, respectively. The real and imaginary part operators are denoted by $\Re\{\cdot\} $ and $\Im\{\cdot\}$, respectively. Moreover, $\diag(\ab)$ or $\diag(\Ab)$ returns a diagonal matrix whose diagonal entries are the same as the elements of $\ab$ or the diagonal entries of $\Ab$. In addition, $\Ab(:,1:J)$ represents the matrix consisting of the left $J$ columns of $\Ab$. Lastly, we use $\Cb_{xy}$ and $\Cb_{x}$ to represent the cross-covariance matrix between $\xb$ and $\yb$ and the auto-covariance matrix of $\xb$, respectively.

\section{System and Quantization Models}\label{sec:system model}
\subsection{System Model}\label{sec:system model_A}
We consider a point-to-point massive MIMO system where a transmitter (Tx) with $\Nt$ antennas communicates with a receiver (Rx) with $\Nr$ antennas. Here, ``massive MIMO'' refers to the deployment of large numbers of antennas at both the Tx and Rx \cite{lin2019transceiver,ning2023beamforming,dai2022delay,qi2022hybrid,zhang2017hybridly,gao2021wideband}. We assume that the Tx is equipped with high-resolution digital-to-analog converters while resolution-adaptive ADCs are deployed at the Rx. The ADC bits at the Rx can be dynamically adjusted to adapt to the CSI. In practice, the resolution-adaptive ADCs can be fabricated by the flash architecture \cite{yoo2002power,nahata2004high,rajashekar2008design}. Let $\ssb \in \Cs^{\Ns}$ ($\Ns\leq \min(\Nt,\Nr) $) be the transmitted signal vector of $\Ns$ data streams. We assume that $\ssb$ follows the Gaussian distribution and $  \Es[\ssb \ssb^\H]=\Ib$. Furthermore, let $\Fb\in\Cs^{\Nt \times \Ns}$ be the precoding matrix with the power constraint $\|\Fb\|_{\Fcl}^2 \leq \Pt$. Here, $\Pt$ denotes the transmit power budget of the Tx. 
The received signal (without quantization) at the Rx can be written as
\begin{equation}\label{eq:received signal}
    \yb
    = \Hb\Fb\ssb + \nb,
\end{equation}
where $\Hb\in \Cs^{\Nr\times \Nt}$ is the channel between the Tx and the Rx, and $\nb$ denotes the additive white Gaussian noise (AWGN) vector, $\nb \sim \Ccl\Ncl(0, \sigman^2 \Ib)$, with $\sigman^2$ being the noise power. 

To characterize the upper bound on the system's performance, we assume the availability of perfect CSI at both the Rx and Tx \cite{bai2013optimization,choi2017resolution,nguyen2020energy,ahmed2017joint, prasad2020optimizing, sheng2020energy,mo2015capacity,ling2019performance,lin2019transceiver,ning2023beamforming,dai2022delay,qi2022hybrid,zhang2017hybridly,gao2021wideband}. The effect of imperfect CSI will be numerically investigated in Section~\ref{sec:imperfect implementation}. We note that the channel can be assumed quasi-static in some point-to-point scenarios (e.g., wireless backhaul). In addition, channel estimation with adaptive-resolution ADCs was studied in \cite{wang2022channel}. Furthermore, an ASIC receiver integrating both the resolution-adaptive ADCs and the channel estimation module was implemented in \cite{castaneda2021resolution}.

\subsection{Channel Model}\label{sec:channel model}
The mmWave propagation environment can be well characterized by the Saleh-Valenzuel channel model \cite{rangan2014millimeter}. Assuming a uniform planar array (UPA) at both the Tx and the Rx, the channel matrix is expressed as \cite{yu2016alternating,lin2019transceiver,qi2022hybrid,zhang2017hybridly}
\begin{equation}\label{eq:SV channel model}
    \Hb=\sqrt{\frac{\Nt\Nr}{\Ncls \Nray}} \sum\limits_{i=1}^{\Ncls}\sum\limits_{l=1}^{\Nray} \alpha_{il} \ab_{\rm r}\left( \theta_{il}^{\rm r},\phi_{il}^{\rm r}\right)\ab_{\rm t}^\H\left(\theta_{il}^{\rm t},\phi_{il}^{\rm t} \right),
\end{equation}
where $\Ncls$ and $\Nray$ indicate the number of clusters and distinct rays within each cluster, respectively, and $\alpha_{il}$ denotes the gain of $l$-th ray in the $i$-th propagation cluster. In addition, $\ab_{\rm r}\left( \theta_{il}^{\rm r},\phi_{il}^{\rm r}\right)$ and $\ab_{\rm t}\left( \theta_{il}^{\rm t},\phi_{il}^{\rm t}\right)$ represent the receive and transmit array response vectors, respectively, where  $\theta_{il}^r$ ($\phi_{il}^{\rm r}$) and $\theta_{il}^\T$ ($\phi_{il}^{\rm t}$) stand for the azimuth (elevation) angles of arrival/departure (AoAs/AoDs) of the $l$-th ray in the $i$-th propagation cluster, respectively. Assume that the Rx deploys a UPA of size $N_{\rm r,h} \times N_{\rm r,v}$ with $N_{\rm r}=N_{\rm r,h} N_{\rm r,v}$. Defining $\rho_{il}^{\rm r}=  \sin(\theta_{il}^{\rm r})\sin(\phi_{il}^{\rm r})$ and $\varrho_{il}^{\rm r} = \cos(\theta_{il}^{\rm r}) $, the array response vector $\ab_{\rm r}\left( \theta_{il}^{\rm r},\phi_{il}^{\rm r}\right)$ at the Rx can be expressed as\cite{yu2016alternating,lin2019transceiver}
\begin{equation}\label{eq:array steer vector for MIMO}
 \begin{aligned}
    \ab_{\rm r}\left( \rho_{il}^{\rm r},\varrho_{il}^{\rm r}\right)
    =\ab_{\rm r,h}\left( \rho_{il}^{\rm r}\right) \otimes \ab_{\rm r,v}\left( \varrho_{il}^{\rm r}\right),
 \end{aligned}
\end{equation}
with
\begin{equation}
\begin{aligned}
      &\ab_{\rm r,h}\left( \rho_{il}^{\rm r}\right)= \frac{1}{\sqrt{N_{\rm r,h}}}[1,e^{j\pi\rho_{il}^{\rm r}}, \cdots,
      e^{j(N_{\rm r,h}-1)\pi\rho_{il}^{\rm r}}]^\T,  \\
      &\ab_{\rm r,v}\left( \varrho_{il}^{\rm r}\right)= \frac{1}{\sqrt{N_{\rm r,v}}}[1,e^{j\pi\varrho_{il}^{\rm r}}, \cdots,
 e^{j(N_{\rm r,v}-1)\pi\varrho_{il}^{\rm r}}]^\T.
\end{aligned}
\end{equation}
The array response vector $ \ab_{\rm t}\left( \theta_{il}^{\rm t},\phi_{il}^{\rm t}\right)$ at the Tx can be modeled similarly.

\subsection{Signal Model with Quantization}
A scalar quantizer is fully characterized by its codebook $\Ccl$ and threshold set $\Tcl$. For a $b$-bit quantizer, we have $\Ccl=\{c_0, \ldots, c_{N_{\rm q}-1}\}$ and $\Tcl=\{ t_0, \ldots, t_{N_{\rm q}} \}$, where $N_{\rm q}=2^b$ is the number of representation levels of the quantizer. Here, we assume $t_0=-\infty$ and $t_{N_{\rm q}}=\infty$, which allows inputs with arbitrary power.\footnote{In practice, the input signal of ADCs outside the range $[t_1,t_{N_{\rm q}-1}]$ can be clipped into the range of $[t_1-\iota,t_{N_{\rm q}-1}+\iota]$ where $\iota$ is an adjustable parameter depending on the constraints of hardware components, e.g., the automatic gain control (AGC). } Let $Q(\cdot)$ denote the quantization function associated with $\Ccl$ and $\Tcl$. For a complex signal $x$, we have $Q(x)=Q(\Re\{x\})+j Q(\Im\{x\})$, with $Q(\Re\{x\})=c_i, i\in \{0,\ldots, N_{\rm q}-1\}$ for $\Re\{x\}\in [t_i, t_{i+1}]$. $Q(\Im\{x\})$ is obtained similarly. 

The Bussgang decomposition can be applied to a vector space in the complex domain  \cite{demir2020bussgang}. Specifically, let $\Qb:\Cs^N\rightarrow \Cs^N$ denote a scalar quantization function and $\zb$ be the quantized output of $\yb$ in \eqref{eq:received signal}. We can write $\zb=\Qb(\yb)$ or equivalently $z_i=Q_i(y_i),\;\forall i$, where $z_i$ and $y_i$ denote the $i$-th element of $\zb$ and $\yb$, respectively; $Q_i(\cdot)$ represents the associated quantization function. For the circularly-symmetric Gaussian random vector $\yb$, the Bussgang decomposition implies
 \begin{equation}\label{eq:Bussgang decomposition}
 \zb=\Qb(\yb)=\Gb \yb +\etab,
 \end{equation}
  where $\Gb= \Cb_{zy}\Cb_{y}^{-1}$ denotes the Bussgang gain, and the distortion term $\etab$ is uncorrelated to $\yb$. In \eqref{eq:Bussgang decomposition}, $\etab$ represents the QD vector with its covariance matrix given by
 \begin{equation}\label{eq:crrl of error term}
     \Cb_{\eta}=\Es[(\zb-\Gb\yb)(\zb-\Gb\yb)^\H]=\Cb_{z}-\Gb\Cb_{yz}.
 \end{equation}
 Furthermore, under some mild assumptions, the Bussgang gain $\Gb$ is shown to be diagonal, as detailed below.

\begin{lemma}[\!\!\cite{jacobsson2017quantized,bjornson2018hardware,demir2020bussgang}]\label{lemma:element-wise quantization}
Consider a circularly-symmetric Gaussian random vector $\yb$ fed into scalar quantizers. With \eqref{eq:Bussgang decomposition} modeling the quantization, we have $\Gb=\diag\left(\gb \right)$, where $g_i=\frac{\Es[Q_i(y_i)y_i^*]}{\Es[|y_i|^2]}$ is the $i$-th element of $\gb$.
 \end{lemma}

 

Substituting \eqref{eq:received signal} into \eqref{eq:Bussgang decomposition}, we obtain the quantized version of the signal received, expressed as 
\begin{equation}\label{eq:quantized received signal}
	\zb=\Gb\Hb\Fb\ssb+\eb,
\end{equation}
where $\eb = \Gb\nb +\etab$ represents the effective noise with covariance matrix $\Cb_e = \Es[\eb \eb^\H]=\Cb_{\eta}+ \sigman^2\Gb^2$. The post-combined signal at the Rx is expressed as
\begin{equation}\label{eq:estimated signal}
    \hat{\ssb}=\Ub^\H\zb=\Ub^\H\Gb\Hb\Fb\ssb+\Ub^\H\eb,
\end{equation}
where $\Ub\in \Cs^{\Nr \times \Nt}$ denotes the combining matrix. Although $\ssb$ is Gaussian distributed, $\eb$ does not follow a Gaussian distribution because of the nonlinear QD. However, we can treat the effective noise vector $\eb$ as a Gaussian random variable and obtain a lower bound of the SE as \cite{hassibi2003much}
\begin{equation}\label{eq:achievable lb}
  \hspace{-3mm} R=\log \det \left( \Ib +  (\Ub^\H\Cb_e\Ub)^{-1}\Ub^\H \Gb \Hb \Fb \Fb^\H\Hb^\H \Gb \Ub\right).
\end{equation}

It is observed that the Bussgang gain $\Gb$ and the QD covariance matrix $\Cb_{\eta}$ are necessary for further analysis and optimization of the SE. For one-bit quantization, closed-form expressions for $\Gb$ and $\Cb_{\eta}$ can be derived based on the arcsine law. However, obtaining those for multi-bit quantization is significantly more challenging. A closed-form expression of $\Gb$ and a diagonal approximation of $\Cb_{\eta}$ were developed in \cite{mezghani2012capacity,bai2013optimization} under the assumption that the quantizer satisfies the following properties:
\begin{align}
    & \Es[z_i-y_i]=0, \\
    & \Es[(z_i-y_i)z_i]=0.
\end{align}
 However, the validity of these assumptions remains unclear, and thus the applicability of these results to general signal distributions and quantizers is uncertain. In the next section, we derive the BAQNM and diagonal approximation from a new perspective, aiming to clarify this uncertainty.


\section{Relations between Optimal Quantization and BAQNM} \label{sec:QD approx}

In this section, we first identify the fundamental properties of optimal quantizers. 
Then we leverage them to obtain the BAQNM and the approximation of the QD covariance matrix. Furthermore, we elaborate on the nuances between applying the BAQNM and the arcsine law to one-bit quantization.

\subsection{Properties of Optimal Quantizers}\label{sec:Properties of Optimal Quantizers}

We first recall the definition of the optimal quantizer \cite{max1960quantizing} below.
\begin{definition}[\!\!\cite{max1960quantizing}]
    Consider a real-valued random variable $X$. Let $f_X(x)$ denote its probability density function (PDF), and let $Q(x)=c_i, i\in \{0,\ldots, N_{\rm q}-1\}$ be its quantized approximation for $x\in (t_i, t_{i+1}]$. The mean square error (MSE) for the quantization can be expressed as
 \begin{equation}\label{eq:MSE defintion}
     \hspace{-2mm}D=\Es\left[\left(Q(x)-x\right)^2\right]=\sum\limits_{i=0}^{N_{\rm q}-1} \int_{t_i}^{t_{i+1}} (x-c_i)^2 f_X(x) {\rm d}x.
 \end{equation}
The optimal quantizer satisfies that its codebook and threshold set, i.e., $\{\Ccl, \Tcl\}$, minimizes $D$.
\end{definition}

Setting the derivatives of $D$ with respect to $t_j$ and $c_j$ to zeros yields
 \begin{align}
     &t_j=\frac{c_j+c_{j-1}}{2},\label{eq:nearest neighbor condition}\\
     &c_j=\frac{\int_{t_j}^{t_{j+1}} x f_X(x) {\rm d}x}{\int_{t_j}^{t_{j+1}} f_X(x) {\rm d}x}, \label{eq:centriod condition}
 \end{align}
which are referred to as the {\it nearest neighbor condition} and the {\it centroid condition}, respectively, \cite[Ch. 6]{gersho2012vector}. They are necessary for the optimal quantizer, also known as the Lloyd-Max quantizer \cite{max1960quantizing} or the optimal non-uniform quantizer. The latter term reflects the fact that the representation levels of the optimal quantizer generally have non-uniform distribution in the real domain. In contrast, a uniform quantizer maintains equal distances between $c_i$ and $c_{i+1}, \forall i$. With this constraint, the quantizer that minimizes $D$ in \eqref{eq:MSE defintion} is referred to as the optimal uniform quantizer.

\begin{remark}
    The centroid condition requires that the representation level of each interval is its mean value.
Mathematically, it can be written as \cite{gersho2012vector}
\begin{equation}\label{eq:centroid condition another form}
       \Es[X|Q(X)]=Q(X),
\end{equation}
which was used in \cite{fletcher2007robust} as a basic assumption for deriving the model \eqref{eq:linear model 2}. Therefore, the BAQNM is limited to the optimal quantizer.
\end{remark}


For a specific input signal, we can employ the Lloyd-Max algorithm \cite{max1960quantizing} that iteratively updates $\Tcl$ and $\Ccl$ based on \eqref{eq:nearest neighbor condition} and \eqref{eq:centriod condition} to find the optimal quantizer. However, this iterative algorithm results in a high time complexity, especially for high-resolution quantization. We herein present an efficient approach to obtain the optimal quantization for Gaussian signals with the proposition below. 

\begin{proposition}\label{lemma:scaling and distortion invariance}
        Let $X$ be a real-valued, zero-mean, and unit-variance Gaussian random variable, and let $Y=\sigmay X$. Then, we have
     \begin{align}
       &\Qy(Y)  =\sigmay \Qx (X)=\sigmay \Qx \left(\frac{Y}{\sigmay}\right), \label{lemma: scaling law}\\
       &\gamma  = \Es\left[(\Qx(X)-X)^2\right]=\frac{\Es\left[(\Qy(Y)-Y)^2\right]}{\sigmay^2} \label{eq:distortion invaniance} ,
     \end{align}
     where $\Qy(Y)$ and $\Qx(X)$ denote the optimal quantized output of $Y$ and $X$, respectively. 
   \end{proposition}

\IEEEproof  See Appendix~\ref{prof:scaling and distortion invariance}. \qedsymbol

\smallskip

We refer to $\gamma$ as the {\it distortion factor} and the properties in \eqref{lemma: scaling law} and \eqref{eq:distortion invaniance} as the {\it scaling law} and {\it distortion invariance}, respectively. The scaling law enables a convenient way to obtain the optimal quantization for any Gaussian signal with a known variance. For example, we can obtain the optimal element-wise quantization of the received signal vector $\yb$ in \eqref{eq:received signal} with covariance matrix 
\begin{equation}
 \Cb_y =\Es[\yb \yb^\H]=\Hb \Fb \Fb^\H \Hb^\H + \sigman^2 \Ib
\end{equation}
based on the optimal quantizer for the standard Gaussian signal \cite{max1960quantizing}.
Regarding the distortion factor, we note the following property.
\begin{proposition}\label{lemma:optimal quantizer condition}
For a zero-mean complex random variable $X=\Re\{X\}+j \Im\{X\}$ with variance $\sigma_X^2$, assume that $\Re\{X\}$ and $\Im\{X\} $ are independent and identically distributed (i.i.d.) with the same variance $\frac{\sigma_X^2}{2}$ and are independently quantized with two identical Lloyd-Max quantizers $Q(\cdot)$. With $\chi= Q(X)-X$, we obtain
\begin{align}
&\Es[Q(X)]  =\Es[X],\\
&\Es[Q(X)\chi^*]  =0, \label{eq:crrlt of output and distortion}\\
&\gamma  =\frac{\Es[\left|\chi \right|^2]}{\Es[\left|X\right|^2]}=\frac{\Es[\Re\{\chi\}^2]}{\Es[\Re\{X\}^2]}=\frac{\Es[\Im\{\chi\}^2]}{\Es[\Im\{X\}^2]}. \label{eq:gamma real and imag part}
 \end{align}
 \end{proposition}
 
\IEEEproof See Appendix~\ref{prof:optimal quantizer condition}. ~\qedsymbol

\smallskip

Propositions~\ref{lemma:scaling and distortion invariance} and \ref{lemma:optimal quantizer condition} reveal fundamental properties of the optimal quantization of Gaussian signals, which had not been previously introduced in the literature. Note that the optimal quantization implies the optimal non-uniform quantizer. However, Proposition~\ref{lemma:scaling and distortion invariance} also holds for the optimal uniform quantizer. The same cannot be concluded for Proposition~\ref{lemma:optimal quantizer condition} because it is derived based on the centroid condition \eqref{eq:centriod condition}, which is generally not satisfied by a uniform quantizer. Nonetheless, we will later numerically verify that Proposition~\ref{lemma:optimal quantizer condition} still approximately holds for the optimal uniform quantizer. Next, we derive the BAQNM and the approximation of the QD covariance matrix based on Propositions~\ref{lemma:scaling and distortion invariance} and~\ref{lemma:optimal quantizer condition}.
 
\subsection{BAQNM and Approximation of the QD Covariance Matrix}\label{sec:BAQNM and QD approximation}

Building on Lemma~\ref{lemma:element-wise quantization} and Proposition~\ref{lemma:optimal quantizer condition}, we can obtain a closed-form expression for the Bussgang gain matrix $\Gb$, as detailed below.
 \begin{corollary}\label{eq:Bussgang gain}
     For a zero-mean complex Gaussian signal vector $\yb=[y_1,\ldots,y_N]^\T \in \Cs^{N}$, assume that the real and imaginary parts of $y_i$ have the same variance and are independently quantized with two identical Lloyd-Max quantizers $Q_i(\cdot)$. Define $\qb=[q_1,\ldots,q_N]^\T$ with $q_i=Q_i(y_i)-y_i, \forall i$ being the quantization error. 
     The Bussgang gain matrix $\Gb$ is given by
     \begin{equation}\label{eq:Bussgang gain matrix}
     	\Gb=\Ib-\Gammab,
     \end{equation}
      where $\Gammab=\diag(\gamma_1,\ldots,\gamma_N) $ with $\gamma_i=\frac{\Es[|q_i|^2]}{\Es[|y_i|^2]}$ being the distortion factor of the $i$-th pair of quantizers.
 \end{corollary}
\IEEEproof With Lemma~\ref{lemma:element-wise quantization}, we have $\Gb=\diag\left(g_1,\ldots,g_N \right)$ and $g_i=\frac{\Es[Q_i(y_i)y_i^*]}{\Es[|y_i|^2]}, \forall i$. Therefore, we have
\begin{align}
     g_i&=\frac{\Es[Q_i(y_i)y_i^*]}{\Es[|y_i|^2]}=\frac{\Es[\left(y_i+q_i \right)y_i^*]}{\Es[|y_i|^2]} \nonumber \\
     &\overset{(d)}{=}1+\frac{\Es[q_i\left(y_i-Q_i(y_i)\right)^*]}{\Es[|y_i|^2]} =1-\frac{\Es[|q_i|^2]}{\Es[|y_i|^2]}\overset{(e)}{=}1-\gamma_i,
\end{align}
where $(d)$ and $(e)$ follow from \eqref{eq:crrlt of output and distortion} and \eqref{eq:gamma real and imag part}, respectively.~\qedsymbol

 \smallskip

Recall that the distortion factor does not depend on the signal variance but only on the resolutions of the quantizers, as shown in Proposition~\ref{lemma:scaling and distortion invariance}. Therefore, once the quantizer resolutions across the RF chains have been determined, we can find $\Gb$ for Gaussian signals undergoing the optimal quantization. The value of the distortion factor for $b \in \{1, \ldots, 5\}$ can be found in \cite{max1960quantizing}. For the optimal quantizer with more than 5 bits, it was shown that its distortion factor can be approximated as \cite[Ch. 6]{gersho2012vector}
 \begin{equation}\label{eq:high resolution gamma approx}
     \gamma(b)\approx\frac{\sqrt{3}\pi}{2}2^{-2b},
 \end{equation}
 where we omit the subscript $i$ without loss of generality and explicitly express $\gamma$ as a function of $b$ for clarity. Equation \eqref{eq:high resolution gamma approx} provides a good approximation of the distortion factor for the high-resolution quantization. However, it incurs large approximation errors for fewer quantization bits. We herein present an approximation that is also valid for low-resolution cases. Specifically, the distortion factors of both the Lloyd-Max and the optimal uniform quantizer can be approximated as
        \begin{align}\label{eq:distortion approx proposed}
            \gamma(b)\approx 2^{-1.74b+0.28}.
        \end{align}


 \begin{figure}[t]
\small
    \centering
    \hspace{-5mm}
    \subfigure[]
    {\label{fig:Distortion factor approx}\includegraphics[width=0.25\textwidth]{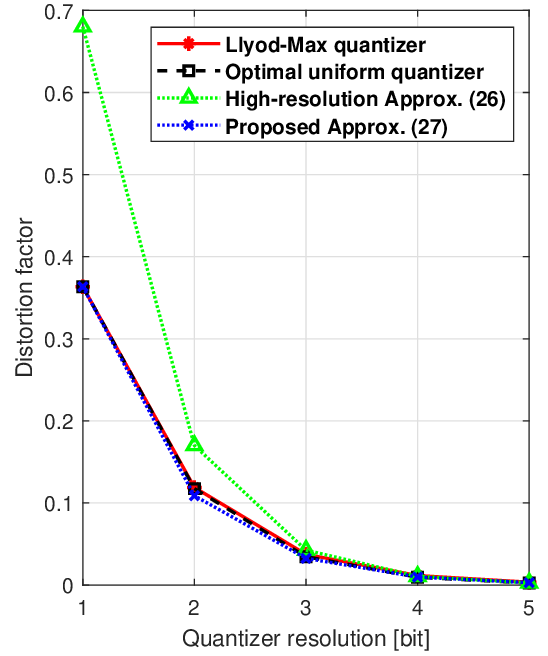}} \hspace{-3mm}
    \subfigure[]
    {\label{fig:model mismatch} \includegraphics[width=0.25\textwidth]{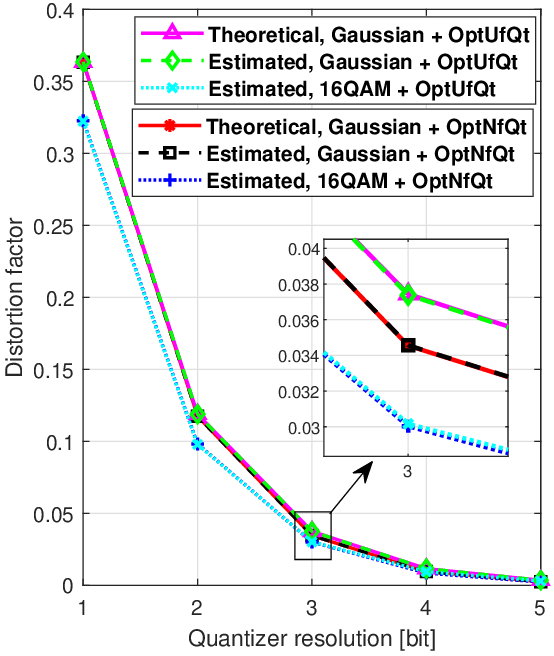}}
    \vspace{-2mm}
    \caption{Distortion factor versus the quantizer resolution $b$. Fig. (a) shows the accuracy of the approximated distortion factor. Fig. (b) shows the accuracy of the estimated distortion factor in an example with $\Nr=\Nt=16$ and $\Ns=4$.}
    \label{fig:Distortion factor vs. resolution}
\end{figure}
Fig.~\ref{fig:Distortion factor approx} shows the approximated distortion factors by \eqref{eq:high resolution gamma approx} and \eqref{eq:distortion approx proposed} compared with the accurate ones of the Lloyd-Max quantizer and the optimal uniform quantizer. It is observed that the distortion factors of the Lloyd-Max quantizer and optimal uniform quantizer are comparable. The proposed approximation in \eqref{eq:distortion approx proposed} can well approach the accurate value of the distortion factors, while the approximation in \eqref{eq:high resolution gamma approx} becomes increasingly inaccurate for fewer bits. For more than five bits, both approximations closely match the theoretical value of the distortion factor, which is omitted from the figure. Note that, compared to \eqref{eq:distortion approx proposed}, the overestimation of the distortion factor by \eqref{eq:high resolution gamma approx} can lead to a significant performance overestimation, as will be numerically verified in Section~\ref{sec:simulation}.

With Corollary~\ref{eq:Bussgang gain}, \eqref{eq:Bussgang decomposition} can be recast as
\begin{equation}\label{eq:BAQNM}
     \zb=(\Ib-\Gammab)\yb + \etab,
\end{equation}
which is exactly the vector form of the BAQNM. Furthermore, with \eqref{eq:BAQNM}, we obtain
\begin{equation}
     \etab=\zb-(\Ib-\Gammab)\yb=\qb+\Gammab\yb,
 \end{equation}
which yields
 \begin{align}
     &    \Cb_{\eta}=\Es[(\qb+\Gammab\yb)(\qb+\Gammab\yb)^\H]=\Cb_{q}-\Gammab\Cb_{y}\Gammab, \label{eq:CoV of error term with q}\\
     &\Cb_{z}=\Cb_{q}+(\Ib-\Gammab)\Cb_{y}-\Cb_{y}\Gammab \label{eq:Cov of z with q},
 \end{align}
 where $\Cb_{q}$ and $\Cb_{z}$ denote the covariance matrices of $\qb$ and $\zb$, respectively.
 As the covariance matrix $\Cb_{y}$ can usually be estimated as a prior, we can obtain closed-form expressions for $\Cb_{\eta}$ and $\Cb_{z} $ by approximating $\Cb_{q}$. The details are presented below.
 \begin{corollary}\label{lemma:approximated Cov of eta and z}
     For a circularly-symmetric Gaussian vector $\yb$, assume that the Lloyd-Max quantizers are adopted for each element of $\yb$. 
     The following approximations hold:
 \begin{align}
     &\Cb_{q} \approx \Gammab\Cb_{y}\Gammab+(\Ib-\Gammab)\diag(\Cb_{y})\Gammab,\\
     &\Cb_{\eta}\approx \Gammab \, \diag(\Cb_{y})(\Ib-\Gammab), \label{eq:diagonal approximation of ComQuad}\\
 & \Cb_{z}\approx \left[\diag(\Cb_{y})\Gammab +(\Ib-\Gammab)\Cb_{y}\right](\Ib-\Gammab).
 \end{align}
 The diagonal entries of $\Cb_{\eta}$ are accurate. The approximation of $\Cb_{\eta}$ is due to neglecting its non-zero off-diagonal entries, and it becomes more accurate with more quantization bits.
 \end{corollary}
 
\IEEEproof See Appendix~\ref{prof:approximated Cov of eta and z}. \qedsymbol

\smallskip


We note that the BAQNM and the approximation of the QD covariance matrix in \eqref{eq:BAQNM} and Corollary~\ref{lemma:approximated Cov of eta and z} coincide with those in \cite{mezghani2012capacity,bai2013optimization}. Furthermore, with \eqref{eq:distortion approx proposed},
 we obtain a more accurate characterization of the BAQNM and closed-form approximations of $\Cb_{q}$, $\Cb_{\eta}$, and $\Cb_{z}$. Unlike \cite{mezghani2012capacity,bai2013optimization}, our analysis unveils that the BAQNM and the diagonal approximation of the QD covariance matrix typically hold for a circularly-symmetric Gaussian random vector quantized with the Lloyd-Max quantizers. Therefore, the BAQNM implies the use of Gaussian signaling and Lloyd-Max quantizers.\footnote{We note that the BAQNM and the diagonal approximation of the QD covariance matrix {\it approximately} hold for Gaussian signals fed into the optimal uniform quantizers. This is because Proposition~\ref{lemma:optimal quantizer condition} approximately holds for the optimal uniform quantizer, as verified in Fig.~\ref{fig:model mismatch}.} Building on this condition, we will perform the joint transmit-receive beamforming design and bit allocation with \eqref{eq:Bussgang gain matrix} and \eqref{eq:diagonal approximation of ComQuad} in Section~\ref{sec:transceiver design}. Furthermore, it is worth noting that without the condition, the system's performance characterized based on the BAQNM and the diagonal approximation of $\Cb_{\eta}$ becomes less accurate. 

As an example, Fig.~\ref{fig:model mismatch} shows the simulated distortion factors for Gaussian signaling and signaling of 16-quadrature amplitude modulation (16-QAM) in comparison with their theoretical values. Both types of the received signals are quantized with the optimal non-uniform quantizer (OptNfQt) and the optimal uniform quantizer (OptUfQt) \cite{max1960quantizing} at the Rx. The simulated distortion factor is obtained as $\frac{1}{I}\sum_{i=1}^I \frac{|s^{(i)}-s_q^{(i)}|^2}{|s^{(i)}|^2}$ where $s^{(i)}$ and $s_q^{(i)}$ denote the $i$-th received signal sample and the quantized one, with $I=10^5$. It is seen that the estimated distortion factors for Gaussian signaling align well with their theoretical values, while those for 16-QAM signaling yield smaller values due to the mismatch between the signal distribution and the quantizers. Such a mismatch also renders the diagonal approximation of the QD covariance matrix less accurate. By identifying the underlying condition, proposing a more accurate closed-form approximation of the distortion factor, and introducing a more efficient method for evaluating the QD covariance matrix, we obtain a more accurate characterization of the BAQNM than the conventional one \cite{mezghani2012capacity,bai2013optimization}. The accuracy improvement can lead to a significantly more reliable performance evaluation, as will be numerically verified in Section~\ref{sec:simulation}.

\vspace{-2mm}
\subsection{One-Bit Case}
The above discussion and the results in Corollary~\ref{lemma:approximated Cov of eta and z} are also valid for the one-bit quantization. Thus, the closed-form expressions of $\Gb$ and $\Cb_{\eta}$ and their connection to the arcsine law can be shown. We first recall the widely used results for one-bit quantization next.
 \begin{lemma}[\!\!\cite{usman2016mmse,li2017channel,atzeni2021channel}]\label{eq:1bit quantizer}
 	Denote the one-bit quantization function as $Q(\yb)=\sqrt{\frac{\beta}{2}}\left[\sgn\left(\Re\{\yb\} \right) + j\sgn\left(\Im\{\yb\} \right) \right]$, where $\sgn(\cdot)$ returns the signs of the real and imaginary parts of each element of $\yb$. Let $\zb= Q(\yb)$. The following equality holds:
 	\begin{align}
 		&\Cb_{zy}  =\sqrt{\frac{2\beta}{\pi}} \Kb^{-\frac{1}{2}} \Cb_y, \\
 		&\Cb_z  =\frac{2\beta}{\pi}\arcsin\left(\Kb^{-\frac{1}{2}}\Cb_y\Kb^{-\frac{1}{2}} \right),\label{eq:Cov_z 1-bit quantizer}\\
 		&\diag\left(\Cb_z \right)  =\beta \Ib,
 	\end{align}
 	where $\Kb=\diag(\Cb_y)$, and the arcsine function is element-wise applied to its matrix argument.
 \end{lemma}


 We note that the quantization in Lemma~\ref{eq:1bit quantizer} is generally not optimal because all the elements of the signal vector $\yb$, even with different variances, yield the same representation levels after quantization. In contrast, the quantization in Corollary~\ref{lemma:approximated Cov of eta and z} indicates that elements of $\yb$ are optimally quantized. However, the results of Corollary~\ref{lemma:approximated Cov of eta and z} and Lemma~\ref{eq:1bit quantizer} coincide in some circumstances. For example, define a general one-bit quantizer as
 \begin{equation}
     Q(x)=\begin{cases}
  c& \textrm{if}~x\geq 0, \\
  -c& \textrm{if}~x<0.
 \end{cases}
 \end{equation}
For a complex random vector $\yb \sim \Ccl\Ncl(\mathbf{0},\sigma_Y^2\Ib)$, we can obtain the optimal one-bit quantizer as $c=\sqrt{\frac{2}{\pi}}\sigma_Y$ based on conditions \eqref{eq:nearest neighbor condition} and \eqref{eq:centriod condition}. 
  According to \eqref{lemma: scaling law}, the optimal one-bit quantized output can be written as
  \begin{equation}
     \hspace{-1mm} Q(\yb)= \frac{\sigma_Y}{\sqrt{\pi}} \left( \! \sgn\left(\frac{\sqrt{2}}{\sigma_Y}\Re\{\yb\}\right) \! + \! j  \sgn\left(\frac{\sqrt{2}}{\sigma_Y}\Im\{\yb\}\right) \! \right).
  \end{equation}
 Because $\Cb_y=\sigma_Y^2\Ib$ has identical diagonal entries, $\Cb_{\eta}$ is also a diagonal matrix \cite{bjornson2018hardware}. Therefore, Corollary~\ref{lemma:approximated Cov of eta and z} yields {\bf accurate} $\Cb_{\eta}$. Based on \eqref{eq:Bussgang gain matrix} and \eqref{eq:diagonal approximation of ComQuad}, we have
 \begin{equation}\label{eq:B and Cb_eta for 1-bit approx expression special case}
     \Gb=0.6366\Ib, \quad \Cb_{\eta}=0.2313\sigma^2_Y\Ib.
 \end{equation}
On the other hand, by setting $\sqrt{\frac{\beta}{2}}=\sqrt{\frac{2}{\pi}}\sigma_Y$, i.e., $\beta=\frac{2}{\pi}\sigma_Y^2$, we can obtain the optimal quantization in Lemma~\ref{eq:1bit quantizer}. 
With $\zb=\Gb\yb+\etab$ modeling the one-bit quantization and in comparison with Lemma~\ref{eq:1bit quantizer}, we obtain
 \begin{align}
 	&\Gb=\sqrt{\frac{2\beta}{\pi}} \Kb^{-\frac{1}{2}}, \label{eq:B for 1-bit closed-form expression}\\
 	& \Cb_{\eta}=\frac{2\beta}{\pi}\arcsin\left(\Kb^{-\frac{1}{2}}\Cb_y\Kb^{-\frac{1}{2}} \right)-\frac{2\beta}{\pi}\Kb^{-\frac{1}{2}}\Cb_y\Kb^{-\frac{1}{2}}. \label{eq:Cb_eta for 1-bit closed-form expression}
 \end{align}
Based on \eqref{eq:B for 1-bit closed-form expression} and \eqref{eq:Cb_eta for 1-bit closed-form expression}, the resulting $\Gb$ and $\Cb_{\eta}$ are the same as those in \eqref{eq:B and Cb_eta for 1-bit approx expression special case}. This alignment justifies our findings in Section~\ref{sec:BAQNM and QD approximation}.



\section{Joint Beamforming and Bit Allocation Design}\label{sec:transceiver design}

We showed in Section~\ref{sec:QD approx} that the BAQNM and the closed-form approximation of the QD covariance matrix hold under the assumption of Gaussian signals undergoing optimal quantization. With this assumption, we can obtain the closed-form expression of the SE based on \eqref{eq:Bussgang gain matrix} and \eqref{eq:diagonal approximation of ComQuad}, which enable us to proceed with the joint design of transmit-receive beamforming and bit allocation in this section. The design problem is formulated next.

\vspace{-2mm}
\subsection{Problem Formulation}
Building on Corollary~\ref{lemma:approximated Cov of eta and z}, the covariance matrix of the QD vector in \eqref{eq:crrl of error term} can be approximated as
\begin{equation}\label{eq:accurate modelling quantization distortion}
    \Cb_{\eta} \approx \Gb\left(\Ib-\Gb\right)\diag(\Cb_y),
\end{equation}
which leads to the following approximation of the covariance matrix of the effective noise vector:
\begin{equation}\label{eq:CoV eff approx}
    \Cb_e \approx \Gb\left(\Ib-\Gb\right)\diag\left(\Hb \Fb \Fb^\H \Hb^\H\right)+ \sigman^2\Gb.
\end{equation}
This approximation becomes more accurate with higher-resolution ADCs. Define $\bb= [b_1,\ldots,b_{\Nr}]$ with $b_i$ being the resolution bit of the $i$-th pair of ADCs. Let $\mathcal{B}=\{1,\ldots,b_{\rm max}\} $ be the set of possible ADC resolutions allocated to an RF chain. Moreover, let $b_{\rm total}$ be the total number of ADC bits available for all RF chains. The joint design of transmit-receive beamforming and bit allocation is formulated as:

\begin{subequations}\label{pb:problem of interest}
    \begin{align}
        \underset{\bb,\Fb,\Ub}{\mathrm{maximize}} & \quad  R \\
        \mathrm{subject~to} & \quad \|\Fb\|^2_{\Fcl}\leq \Pt,\label{constr:tx power budget}\\
         & \quad \sum\limits_{i=1}^{\Nr}b_i=  \left \lfloor \varsigma b_{\rm total} \right \rfloor,  \label{constr:bitsum limit}\\
         & \quad  b_i \in \mathcal{B},~i=1,\ldots,\Nr, \label{constr:integer constraint}
    \end{align}
\end{subequations}
where $R$ is given in \eqref{eq:achievable lb}, and $\varsigma \in (0,1]$ is the fraction of active ADC bits so that the total number of active ADC bits is $\lfloor \varsigma b_{\rm total} \rfloor$. Here, $ \left \lfloor x \right \rfloor $ denotes the nearest integer smaller than $x$. Problem \eqref{pb:problem of interest} has a mixed-integer nature and coupled variables in the objective function, making it challenging to solve. Observing that the design of $\{\Fb, \Ub\}$ depends on $\bb$, we propose a two-stage optimization framework wherein we first solve $\{\Fb, \Ub\}$ with a given $\bb$, and then find an efficient solution to $\bb$ based on the proposed beamforming algorithm. The details are elaborated next.

\subsection{Proposed Solution for \eqref{pb:problem of interest}}

\subsubsection{Beamforming Design}
For a given $\bb$, the Bussgang gain matrix $\Gb$ is fixed according to Corollary~\ref{eq:Bussgang gain}. Therefore, the beamformers $\{\Fb, \Ub\}$ can be obtained by solving the following problem:
\begin{equation}\label{pb:subproblem of Fb and Ub}
        \underset{\Fb,\Ub}{\mathrm{maximize}} \;  R,\ \mathrm{subject~to} \; \eqref{constr:tx power budget},
\end{equation}
which is non-convex. To address the challenge, we propose to solve an equivalent but more tractable weighted MSE minimization problem. The MSE matrix of the post-combined signals at the Rx is given by
\begin{align}\label{eq:MSE matrix DBF}
  \hspace{-2mm} \Eb =  \Es\left[\left(\hat{\ssb}-\ssb \right)\left(\hat{\ssb}-\ssb \right)^\H \right] & =\Ub^\H\left(\Gb\Hb\Fb\Fb^\H \Hb^\H \Gb + \Cb_e \right)\Ub +\Ib \notag \\
  & \phantom{=} \ -\Ub^\H \Gb \Hb\Fb -\Fb^\H\Hb^\H \Gb \Ub,
\end{align}
where $\Cb_e$ is given in \eqref{eq:CoV eff approx}. The weighted MSE minimization problem is written as
 \begin{align}\label{pb:precoder design WMMSE}
      \underset{\Ub, \Fb,\Wb}{\mathrm{minimize}} & \quad f(\Ub, \Fb,\Wb)= \tr\left( \Wb \Eb\right)-\log \det\left( \Wb\right) \\
     \mathrm{subject~to} &  \quad  \eqref{constr:tx power budget},\nonumber
 \end{align}
where $\Wb \succeq 0$ is an introduced weighted matrix. The following proposition establishes the equivalence of this problem to \eqref{pb:subproblem of Fb and Ub}.
\begin{proposition}\label{prop:WMMSE equivalence DBF}
 The optimal solutions to $\Wb$ and $\Ub$ for \eqref{pb:precoder design WMMSE} are given by
\begin{align}
&\Wb = \Ib + \Fb^\H \Hb^\H\Gb \Cb_e^{-1} \Gb \Hb \Fb, \label{eq:solution to W refined} \\
&\Ub =\left(\Gb\Hb\Fb\Fb^\H \Hb^\H \Gb + \Cb_e \right)^{-1}\Gb\Hb \Fb \label{eq:solution to U}.
\end{align}
It can be shown that the optimal solutions to $\{\Ub,\Fb\}$ for \eqref{pb:precoder design WMMSE} are the same as those for \eqref{pb:subproblem of Fb and Ub}.
\end{proposition}

\IEEEproof See Appendix~\ref{sec:appendices}.\qedsymbol

\smallskip

Note that $f(\Ub, \Fb,\Wb)$ is convex with respect to one variable when the others are fixed. Therefore, we can use an alternating optimization procedure to solve problem \eqref{pb:precoder design WMMSE}. Since the solutions to $\Ub$ and $\Wb$ are given, we delineate the solution to $\Fb$ next. With some algebra, the subproblem of the precoder design can be expressed as
\begin{subequations}\label{eq:precoder design WMMSE}
 \begin{align}
     \underset{\Fb}{\mathrm{minimize}} & \quad  \tr(\Jb \Fb \Fb^\H)  - 2 \Re\{\tr\left(\Wb\Ub^\H \Gb \Hb \Fb \right)\}\\
     \mathrm{subject~to} & \quad  \tr(\Fb \Fb^\H)\leq \Pt,
 \end{align}
\end{subequations}
where $\Jb = \Hb^\H \left(\Gb\Ub \Wb \Ub^\H  + \diag \left(\Ub \Wb \Ub^\H \right)(\Ib-\Gb)\right)\Gb\Hb$. Problem \eqref{eq:precoder design WMMSE} is convex and admits a closed-form solution. Specifically, we first obtain the Lagrangian function as
\begin{align}
    L(\Fb,\mu)&= \tr(\Jb \Fb \Fb^\H) - 2 \Re\{\tr\left(\Wb\Ub^\H \Gb \Hb \Fb \right)\} \nonumber \\
    & \phantom{=} \ + \mu(\tr(\Fb \Fb^\H)- \Pt),
\end{align}
where $\mu \geq 0$ is the Lagrangian multiplier. Leveraging the first-order condition of optimality, we obtain
\begin{equation}\label{eq:Fb closed-form}
    \Fb=\left( \Jb + \mu \Ib\right)^{-1}\Hb^\H \Gb \Ub \Wb,
\end{equation}
where $\mu$ satisfies $\mu\left(\|\Fb\|^2_{\Fcl}-\Pt \right)=0$ and can be obtained via the bisection search in the interval $\left[0, \frac{\|  \Hb^\H\Gb\Ub \Wb\|_{\Fcl}}{\sqrt{\Pt}}\right]$.

The alternating optimization procedure for updating $\Fb$ and $\Ub$, referred to as AltMin beamforming (AltMin-BF) design, is summarized in Algorithm~\ref{alg:WMMSE-AltMin DBF Design}. Because the alternating updates of $\Wb$, $\Ub$, and $\Fb$ result in a nondecreasing sequence of objective values, which are upper bounded due to the power constraint, the convergence of Algorithm~\ref{alg:WMMSE-AltMin DBF Design} is guaranteed. We initialize $\Fb$ based on the WF method and set $\Wb=\Ib$.

\begin{algorithm}[t]
\small
\caption{AltMin-BF design}\label{alg:WMMSE-AltMin DBF Design}
\LinesNumbered 
\KwOut{$\Fb,\Ub$}
Initialize $\bb,\Fb,\Wb,\varepsilon $.\\
\Repeat{ $\left|\log \det(\Wb')-\log\det(\Wb) \right|\leq  \varepsilon  $}{
$\Wb'\leftarrow \Wb$.\\
$ \Ub \leftarrow \left(\Gb\Hb\Fb\Fb^\H \Hb^\H \Gb + \Cb_e \right)^{-1}\Gb\Hb \Fb$.\\
$ \Wb \leftarrow  \Ib + \Fb^\H \Hb^\H\Gb \Cb_e^{-1} \Gb \Hb \Fb$.\\
Update $\Fb$ by \eqref{eq:Fb closed-form}.\\
 }
\end{algorithm}

\subsubsection{Bit Allocation}
With $\Fb$ and $\Ub$ obtained by Algorithm~\ref{alg:WMMSE-AltMin DBF Design}, the bit allocation problem is formulated as
\begin{align}\label{pb:subproblem bit allocation}
    \underset{\bb}{\mathrm{maximize}} & \quad R(\bb)\\
    \mathrm{subject~to} & \quad \text{\eqref{constr:bitsum limit} and \eqref{constr:integer constraint}},  \nonumber
\end{align}
where $R(\bb)$ represents the SE achieved with $\bb$. The non-convex integer nature makes problem \eqref{pb:subproblem bit allocation} again challenging to solve. An exhaustive search (ES) can be performed to find the optimal solution. However, it requires excessively high complexity, especially for a large number of antennas. 

To overcome the challenge, we propose a low-complexity greedy pair-order search-based beamforming and bit allocation (GPOS-BFBA) in Algorithm~\ref{alg:GPOS BFBA}. Specifically, for initialization, we first assume that all the ADCs employ $b_{\rm max}$ bits, i.e., $b_i=b_{\rm max}, \forall i$. In steps 2--9, the ADC bits in each RF chain are gradually decreased to one until the total bit requirement is reached. This is based on the fact that more ADC bits offer higher SE. 
In steps 11--17, a neighbor search procedure is performed to update the solution to $\bb$. Specifically, let $\bb^{(\ell)}$ be the candidate in the $\ell$-th iteration. We define the neighbor set of $\bb^{(\ell)}$ as
\begin{align}
    \label{eq_neighborset}
    \hspace{-0.25cm}\Ncl(\bb^{(\ell)}) \! = \! \left\{\! \tilde{\bb}^{(\ell)}\! : \tilde{\bb}^{(\ell)} \!=\! \bb_{[i \leftrightarrow j]}^{(\ell)}\; \text{if} \; b^{(\ell)}_{i} \!\neq b^{(\ell)}_{j}, \tilde{\bb}^{(\ell)} \notin \mathcal{L} \!\right\},
\end{align}
where $\bb_{[i \leftrightarrow j]}^{(\ell)}$ is obtained by swapping the $i$-th and $j$-th elements of $\bb^{(\ell)}$, i.e., 
\begin{equation}
\small
    \bb_{[i \leftrightarrow j]}^{(\ell)} = [ b_1^{(\ell)}, \ldots, b_{i-1}^{(\ell)}, b_{j}^{(\ell)}, b_{i+1}^{(\ell)}, \ldots, b_{j-1}^{(\ell)}, b_{i}^{(\ell)}, b_{j+1}^{(\ell)}, \ldots, b_{N_{\mathrm{r}}}^{(\ell)} ]^\T.
\end{equation}
Furthermore, $\mathcal{L} = \cup_{m = 1}^{\ell - 1} \Ncl(\bb^{(m)})$ is the list of the candidates examined in the previous iterations. A neighbor point should not belong to this list to avoid a cycling search. Let $S$ denote the number of feasible neighbors with $S\leq \frac{1}{2}\Nr(\Nr-1)$. The neighbor set $\Ncl(\bb^{(\ell)})$ can be obtained by randomly choosing $\Nb$ candidates from the $S$ candidates. For each neighbor point, the beamformers, i.e., $\Fb$ and $\Ub$, are obtained using Algorithm~\ref{alg:WMMSE-AltMin DBF Design}, and the corresponding SE is computed, as in step 13. Then, the best neighbor point $\bb^{(\ell)\star}$ that offers the highest SE is found as in step 14. In step 15, the best solution $\bb^{\star}$ can be updated as $\bb^{(\ell)\star}$ if the latter achieves a higher SE. This iterative process is repeated for $I_2$ iterations or until convergence. Finally, the beamformers and the resolution vector are returned in step~18. The iterative procedure in Algorithm~\ref{alg:GPOS BFBA} jointly solves the resolution vector $\bb$ and beamformer $\{\Fb, \Ub\}$ in each iteration to improve system SE, which guarantees nondecreasing SE over iterations.

\begin{algorithm}[t]
\small
\caption{GPOS-BFBA design}\label{alg:GPOS BFBA}
\LinesNumbered 
\KwOut{$\bb^{\star},\Fb^\star, \Ub^\star$}
 Initialize $\varsigma $, $b_{\rm total}$, $b_{\rm max}$, $\Nb$, and set $b_i=b_{\rm max},\ \forall i$.\\
\For{$n=1,\ldots,\Nr$}{
   \While{$b_n\geq 2$ and $\sum_{i=1}^{\Nr}b_i >\left \lfloor \varsigma b_{\rm total} \right \rfloor$}{
    $b_n=b_n-1$.\\
    }
    \If{$\sum_{i=1}^{\Nr}b_i =\left \lfloor \varsigma b_{\rm total}\right \rfloor$}{
    {break.}
    }
 }
 Set $\bb^\star= [b_1,\ldots,b_{\Nr}]^\T$, $\ell = 1$, $\bb^{(\ell)}=\bb^\star$.\\

 \For{$\ell = 1,\ldots,I_2$}{
    Construct the neighbor set $\Ncl(\bb^{(\ell)} )$ based on \eqref{eq_neighborset}.
                
                
        
    
    Obtain $\Fb$ and $\Ub$ using Algorithm \ref{alg:WMMSE-AltMin DBF Design} and the resultant SE for each neighbor point in $\Ncl(\bb^{(\ell)} )$. 
    
    Set $\bb^{(\ell)\star}$ to the neighbor point that offers the largest SE.

    Update $\bb^{\star} = \bb^{(\ell)\star}$ if $R(\bb^{(\ell)\star}) > R(\bb^{\star})$.

    Set $\bb^{(\ell+1)}=\bb^{\star}$ as the candidate for the next iteration.
 }

    Output $\{\Fb^\star,\Ub^\star \}$ associated with $\bb^{\star}$.
\end{algorithm}

\subsection{Complexity Analysis}\label{sec:complexity analysis}
In massive MIMO systems, we often have $\Ns \ll \min(\Nt,\Nr)$. Thus, the per-iteration complexity of Algorithm~\ref{alg:WMMSE-AltMin DBF Design} can be shown as $\Ocl(3\Nt^3+3\Nr^3+8\Nt^2\Nr+8\Nt\Nr^2)$, which is mainly due to the computation of matrix multiplications, inverses, and determinants. Therefore, the overall complexity of Algorithm~\ref{alg:WMMSE-AltMin DBF Design} is in the order of $I_1\Ocl(3\Nt^3+3\Nr^3+8\Nt^2\Nr+8\Nt\Nr^2)$, where $I_1$ denotes the total number of iterations required. In addition, since the size of $\Ncl(\bb^{(\ell)})$ is $\Nb$, the complexity of Algorithm~\ref{alg:GPOS BFBA} is in the order of $I_2 I_1\Nb\Ocl(3\Nt^3+3\Nr^3+8\Nt^2\Nr+8\Nt\Nr^2)$. Here, $I_2$ is the total number of iterations in the search procedure of Algorithm~\ref{alg:GPOS BFBA}. In contrast, the complexity of the ES method for solving problem \eqref{pb:subproblem bit allocation} is $\bmax^{\Nr}\Ocl(2\Nt^3+2\Nt\Nr^2+2\Nt^2\Nr)$. Compared to ES, Algorithm~\ref{alg:GPOS BFBA} has an enormous reduction in complexity. Furthermore, numerical simulations show that only a few iterations are sufficient for Algorithm \ref{alg:WMMSE-AltMin DBF Design} to obtain the best neighbor point in step 14. With a small value of $I_1$, the complexity of Algorithm~\ref{alg:GPOS BFBA} is comparable to that of Algorithm~\ref{alg:WMMSE-AltMin DBF Design} when $\Nb$ is significantly smaller than the number of Tx/Rx antennas.

\subsection{Implementation Framework}\label{sec:implementation framework}
In practical massive MIMO systems, CSI is typically estimated through pilot training. During this phase, resolution-adaptive ADCs can be configured to use high resolutions, (e.g., $8$--$12$ bits), under which the QD becomes negligible \cite{choi2017resolution}. This enables efficient CSI estimation using well-established techniques, such as linear minimum mean square error (LMMSE) estimator, compressive sensing \cite{berger2010application}, and orthogonal matching pursuit \cite{lee2016channel}. Assuming the CSI is estimated at the receiver, the proposed joint beamforming and bit allocation algorithm can be efficiently executed, given sufficient computational resources. The resulting precoding matrix is then fed back to the transmitter to enable efficient data transmission \cite{chae2008TSP}. This framework significantly reduces the frequency of information exchange between the transmitter and receiver, thus reducing the signaling overhead. 

In this framework, the main performance limitation may arise from the efficiency of feeding back the precoding matrix. The effect of imperfect precoder on the SE will be evaluated numerically in Section~\ref{sec:imperfect implementation}.

\section{Numerical Results and Discussion}\label{sec:simulation}

We herein provide numerical results to demonstrate the performance of the proposed designs. For all numerical simulations, we use the Saleh-Valenzuel channel model \eqref{eq:SV channel model}, which can well characterize mmWave propagation environments \cite{rangan2014millimeter}. In all simulations, we set $\Ncls=5, ~\Nray=8$, and $\alpha_{il}\sim \Ccl\Ncl(0,1), \forall i,l$ \cite{yu2016alternating}. The azimuth (elevation) AoAs and AoDs, i.e., $\theta_{il}^r$ ($\phi_{il}^{\rm r}$) and $\theta_{il}^\T$ ($\phi_{il}^{\rm t}$), follow the Laplacian distribution with uniformly distributed mean angles over $\left(-\pi, \pi \right]$ and $\left(-\frac{\pi}{2}, \frac{\pi}{2} \right]$ with angular spread of $10$ and $3$ degrees, respectively \cite{el2014spatially,akdeniz2014millimeter}.
Furthermore, the SNR is defined as SNR~$= \frac{\Pt}{\sigma_{\rm n}^2}$. The other parameters are detailed in each figure. All reported results are averaged over $10^3$ channel realizations. Furthermore, we use the simulated  $\Cb_{\eta}$ rather than its diagonal approximation as in \eqref{eq:accurate modelling quantization distortion} to evaluate the SE, thanks to the scaling law in Proposition~\ref{lemma:scaling and distortion invariance}. To obtain the simulated $\Cb_{\eta}$, we randomly generate $10^5$ Gaussian signal vectors for transmission and determine the optimal quantization of the received signals based on Proposition~\ref{lemma:scaling and distortion invariance}. This process yields $10^5$ sample variances of the QD. By averaging these sample variances, we obtain the simulated $\Cb_{\eta}$. Note that the simulated  $\Cb_{\eta}$ contains non-zero off-diagonal entries of $\Cb_{\eta}$, enabling more practical performance evaluation compared to using \eqref{eq:accurate modelling quantization distortion}.
 
\vspace{-3mm}
\subsection{Performance Evaluation}\label{sec:performance evalaution}
\begin{figure}[tb]
\small
    \centering
    \vspace{-4mm}
    \hspace{-10mm}
        \subfigure[AltMin-BF algorithm.]
    {\label{fig:Convergence_BF} \includegraphics[width=0.25\textwidth]{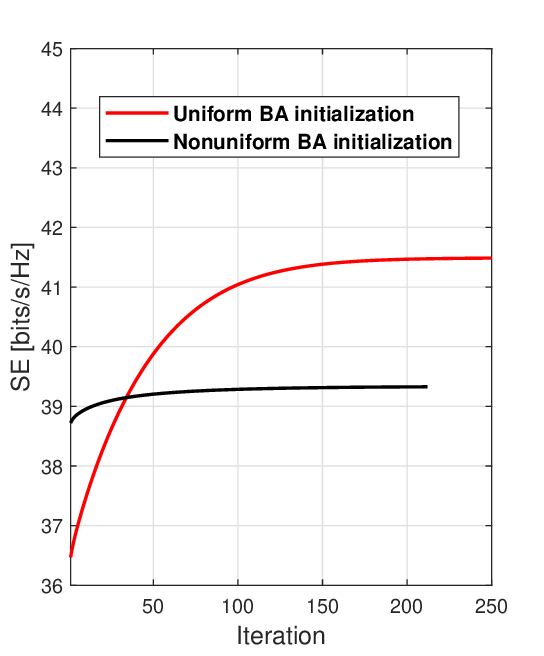}}\hspace{-0.3cm}
            \subfigure[GPOS-BFBA algorithm.]
    {\label{fig:Convergence_GPOS} \includegraphics[width=0.25\textwidth]{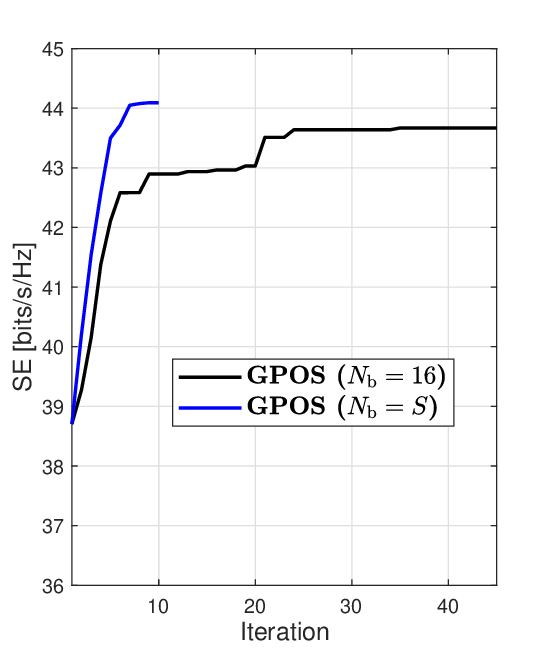}}\hspace{-8mm}
    \caption{Convergence of the proposed algorithms with $\Nt=\Nr=64$, $\Ns=8$,  $\SNR=10~\dB$, $b=2$, $\bmax=8$, and $ \varsigma=1$.}
    \label{fig:convergence}
\end{figure}

Figs.~\ref{fig:Convergence_BF}~and~\ref{fig:Convergence_GPOS} show the convergence of the proposed beamforming design and the GPOS-BFBA algorithm with $\Nr=\Nt=64$, $\Ns=8$, $\SNR =10~\dB$, $b=2$, $\bmax=8$, and $\varsigma=1$. In Fig.~\ref{fig:Convergence_BF}, we show the convergence of the AltMin-BF algorithm with two methods for initializing the bit allocation (BA) , namely the uniform BA and the heuristic non-uniform BA employed in Algorithm~\ref{alg:GPOS BFBA}. With the latter, the AltMin-BF algorithm requires fewer iterations for convergence. This is because the BA in steps 2--9 of Algorithm~\ref{alg:GPOS BFBA} primarily allocates $8$-bit and $1$-bit ADCs to RF chains. These high-resolution ADCs significantly alleviate the impact of the QD, leaving little room for further SE improvement. Therefore, fewer iterations are required for convergence. Fig.~\ref{fig:Convergence_GPOS} shows the convergence of the GPOS-BFBA algorithm with $\Nb\in\{16,S\}$ (represented by ``GPOS ($\Nb=16$)'' and ``GPOS ($\Nb=S$)'', respectively) with the same stopping criterion. Specifically, the algorithm is terminated once the highest SE found remains unchanged over $10$ consecutive iterations. As expected, with a smaller size of the neighbor set, the GPOS-BFBA algorithm converges to a lower SE with the advantage of a lower complexity. We note that more iterations for $\Nb=16$ do not result in higher complexity compared to $\Nb=S$, as significantly fewer neighbor points need to be evaluated in each iteration.

In the subsequent figures, we show the SE and EE performance of the proposed schemes. For comparison, we consider the following baselines:
\begin{enumerate}
    \item The optimal full-resolution scheme, where the eigen-mode beamforming design with WF power allocation is adopted. In this scheme, we set $\Ub=\Zb(:,1:\Ns)$ and $\Fb=\Vb(:,1:\Ns)\Pb^{\frac{1}{2}}$, where $\Zb$ and $\Vb$ are obtained from the singular value decomposition of the channel matrix, i.e., $\Hb=\Zb\Sigmab\Vb^\H$. Here, we have $\Pb= \diag(p_1,\ldots,p_{\Ns})$, where $p_i$ represents the power allocated to the $i$-th data stream and is obtained by the WF method. We refer to this baseline as ``Optimal".
    \item The WF strategy applied to the low-resolution system. Specifically, based on \eqref{eq:quantized received signal}, we can obtain the WF beamformer with the effective channel $\Hb_{\rm eff}=\Gb\Hb$. This scheme is referred to as ``WF".
    \item Random BA strategy combined with the proposed beamforming design. In this baseline, given total active ADC bits $b_{\rm total}$ and $\bmax=8$, we allocate $7$ and $8$ bits randomly to some of the $\Nr$ RF chains, while the remaining ones are assigned to $1$ and $2$ bits. After the BA, the AltMin-BF algorithm is utilized to maximize the SE. We use the term ``Random BA + BFpp" to refer to this scheme in the subsequent discussion.
    \item Proposed beamforming design combined with a genetic algorithm (GA). The combiner and BA are jointly designed based on the GA in \cite{ahmed2017joint}. We herein modify the algorithm to jointly consider both precoder and combiner design. We refer to this baseline as ``BFpp + GA".   
\end{enumerate}
  In all simulations, we set the $b$-bit ADCs for all RF chains for the WF and AltMin-BF algorithms and set $b_{\rm total}=\Nr b$ for all BA schemes. 


\begin{figure*}[t]
\small
    \hspace{-5mm}
    \subfigure[$\Nr=64$ and $b=2$.]
        {\label{fig:SEvsSNR_b2} \includegraphics[width=0.37\textwidth]{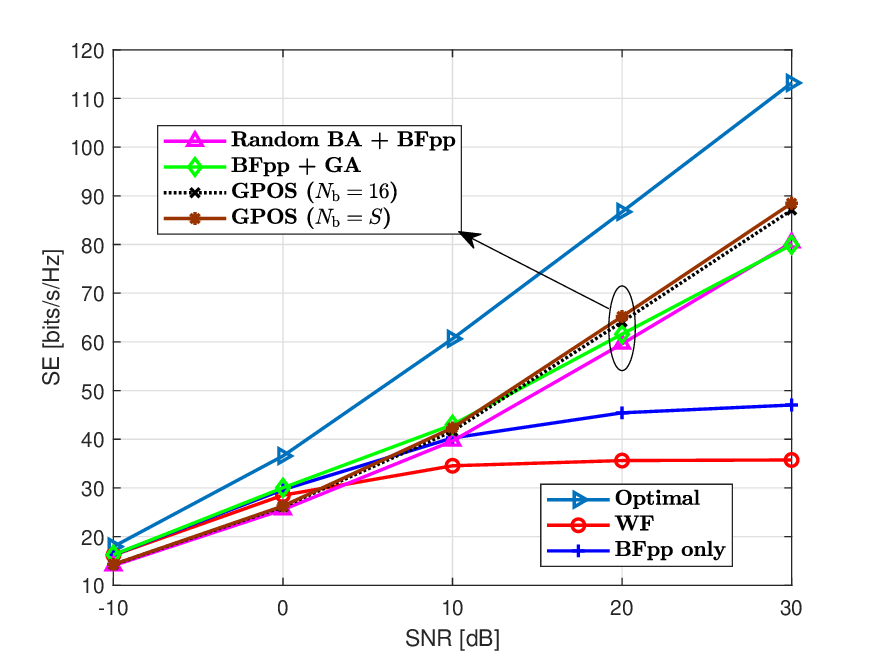}}\hspace{-7mm}
    \subfigure[$\Nr=64$ and $\SNR=30~\dB$.]
        {\label{fig:SEvsBit_SNR30} \includegraphics[width=0.37\textwidth]{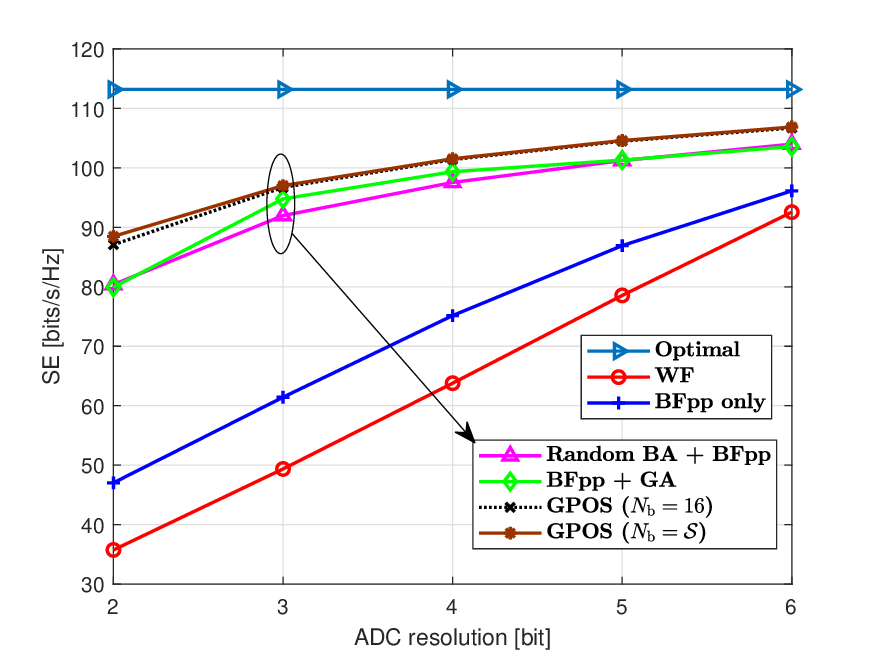}}\hspace{-7mm}
    \subfigure[$b=2$ and $\SNR=30~\dB$.]
    {\label{fig:SEvsNr_SNR30} \includegraphics[width=0.37\textwidth]{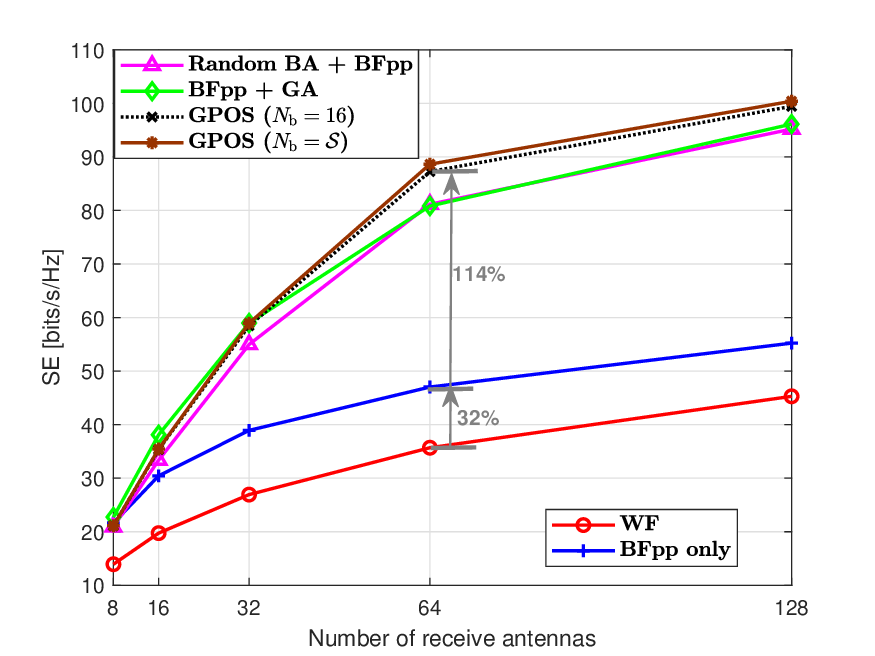}}\hspace{-0.8cm}
    \caption{SE performance with $\Nt=64$, $\Ns=8$, $\bmax=8$, and $\varsigma=1$.}
    \label{fig:Comparision of SE performance versus SNR-Nr-bit evl}
    \vspace{-5mm}
\end{figure*}


  Figs.~\ref{fig:SEvsSNR_b2}--\ref{fig:SEvsNr_SNR30} show the SE of considered schemes versus the SNR, ADC resolution, and the number of receive antennas, respectively, with $\Nt=64$, $\Ns=8$, $\bmax=8$, and $\varsigma=1$. Here, the ``BFpp only'' represents the proposed AltMin-BF algorithm with uniform BA. Based on those figures, we draw the following observations. 
  \begin{itemize}
      \item First, the GPOS-BFBA algorithm, using a neighbor set of size $16$, achieves SE comparable to that obtained with a full neighbor set while significantly reducing complexity. Moreover, the proposed joint beamforming and BA design outperforms the baseline BA schemes, particularly when fewer ADC bits are used per RF chain and at high SNRs. 
      \item Second, the proposed beamforming algorithm significantly outperforms the WF solution. It is observed that the ``BFpp only'' attains $32\%$ SE improvements compared to the ``WF'' scheme with $\Nr=64$, $\SNR=30~\dB$, and $b=2$. 
      \item Finally, the proposed joint beamforming and BA design significantly outperforms the beamforming with uniform BA strategy, especially for large-scale MIMO systems at high SNRs. For example, a $114\%$ SE gain is achieved by the ``GPOS ($\Nb=16$)'' compared to the ``BFpp only'' with $\Nr=64$, $\SNR=30~\dB$, and $b=2$.
  \end{itemize}


\begin{table}[t]
\small
\renewcommand\arraystretch{1.2}
\caption{Comparison of SE gains (\%) achieved by the ``GPOS ($\Nb=16$)'' relative to the WF baseline, based on \eqref{eq:high resolution gamma approx} and \eqref{eq:distortion approx proposed}, with $\SNR=30$~dB.}
\label{tb:Comparison of SE gains}
\centering
    \begin{tabular}{c|ccc} 
    \hline
      $b$ & $2$ & $3$ &  $4$  \\
    \hline
    Conventional approx. \eqref{eq:high resolution gamma approx}&  $\bf 175.85$ & $105.50$ & $62.15$ \\
    \hline
    Proposed approx. \eqref{eq:distortion approx proposed} &  $\bf 143.69$& $95.95$ & $58.99$\\
    \hline
    \end{tabular}
\end{table}

Note that all numerical results in Section~\ref{sec:simulation} are based on the proposed approximation of the distortion factor \eqref{eq:distortion approx proposed}. We perform simulations based on the conventional approximation \eqref{eq:high resolution gamma approx} and show the comparison of SE gains achieved by the ``GPOS ($\Nb=16$)'' relative to the WF baseline in Table~\ref{tb:Comparison of SE gains}. It is seen that the SE gains calculated with the conventional approximation \eqref{eq:high resolution gamma approx} are overestimated compared to those based on our proposed approximation \eqref{eq:distortion approx proposed}, especially with fewer quantization bits. For instance, at $b=2$, the overestimation in SE gain can reach up to $30\%$. This discrepancy arises because the conventional approximation yields a larger distortion factor, which overemphasizes the benefit of QD-aware designs. Therefore, the proposed approximation \eqref{eq:distortion approx proposed} enables more accurate performance evaluation.


\begin{table}[t]
\small
\renewcommand\arraystretch{1.2}
\caption{Average time cost ([s]) with $\Nt=64$, $\Ns=8$, $\bmax=8$, $ \varsigma=1$, $b_{\rm total}=2\Nr$, and $\SNR=30~\dB$.}
\centering
    \begin{tabular}{c|c|c|c} 
    \hline
      $\Nr$  & GPOS ($\Nb=16$) & Random BA  & GA  \\
    \hline
    $64$   & $\bf 12.3$ & $31.5$ &  $212.1$\\
        \hline
    $128$   & $\bf 6.0$ & $35.0$ &  $1201.1$\\
    \hline
    \end{tabular}
\label{tb:TC Comparison}
\end{table}

Table~\ref{tb:TC Comparison} lists the average run time of the considered joint beamforming and BA schemes for $\Nr \in \{64, 128\}$ with $\Nt=64$, $\Ns=8$, $\bmax=8$, $ \varsigma=1$, $b_{\rm total}=2\Nr$, and $\SNR=30~\dB$. The execution time of all the compared schemes is evaluated with the CPU of Xeon Gold 6230. It is seen that the GA-based BA design is the most time-consuming among all schemes, although we employ parallel computing to evaluate the fitness of the populations in each iteration to accelerate the convergence. The proposed GPOS-BFBA design with $\Nb=16$ requires significantly less run time compared to the baselines, demonstrating its efficiency considering its superior SE performance. In addition, the WF scheme has a complexity of $\Ocl(\Nt\Nr\Ns)$, which is lower than that of the proposed schemes at the expense of substantially poorer performance in low-resolution systems, as observed in Fig.~\ref{fig:Comparision of SE performance versus SNR-Nr-bit evl}.



\subsection{Impact of Imperfect Implementation}\label{sec:imperfect implementation}
\begin{figure}[t]
	\small
	\centering	
	\includegraphics[width=0.45\textwidth]{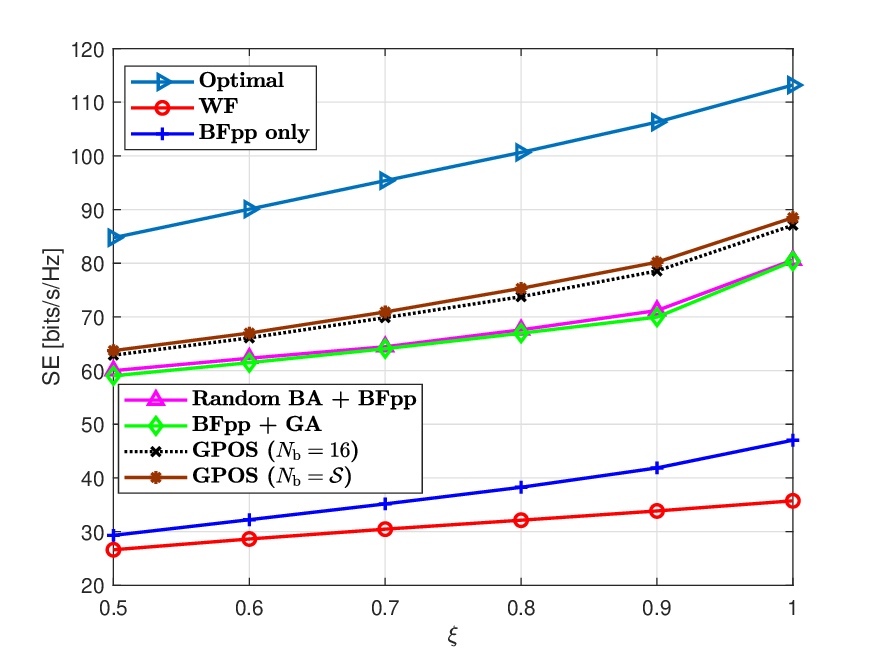}
	\vspace{-2mm}
	\caption{SE performance under imperfect CSI with $\Nt=\Nr=64$, $\Ns=8$, $\textrm{SNR} = 30$~dB, $b=2$, and $\bmax=8$.}
	\label{fig:SE performance versus CSI}
\end{figure}

The proposed designs rely on the perfect CSI and beamforming matrices, which are challenging to obtain in practice. Hence, we evaluate the impact of imperfect CSI and precoder feedback on the SE in the following. The estimated channel matrix (imperfect CSI) $\hat{\Hb}$ can be modeled as $\hat{\Hb}=\xi \Hb + \sqrt{1-\xi^2}\Eb$ \cite{jacobsson2019linear,chu2019efficient}
where $\Hb$ is the true channel, $\xi \in [0,1]$ controls the CSI estimation accuracy, and $\Eb$ represents the estimation error with entries following distribution $\mathcal{CN}(0,1)$. Fig.~\ref{fig:SE performance versus CSI} shows the SE of the considered schemes under imperfect CSI with $\Nt=\Nr=64$, $\Ns=8$, $\textrm{SNR} = 30$~dB, $b=2$, and $\bmax=8$. It is observed that imperfect CSI affects all the proposed and baseline schemes similarly, resulting in comparable levels of SE degradation. Notably, with higher CSI accuracy, the proposed schemes demonstrate more significant SE gains over the baselines, further validating their efficiency.

\begin{figure}[tb]
	\small
	\centering	
	\includegraphics[width=0.45\textwidth]{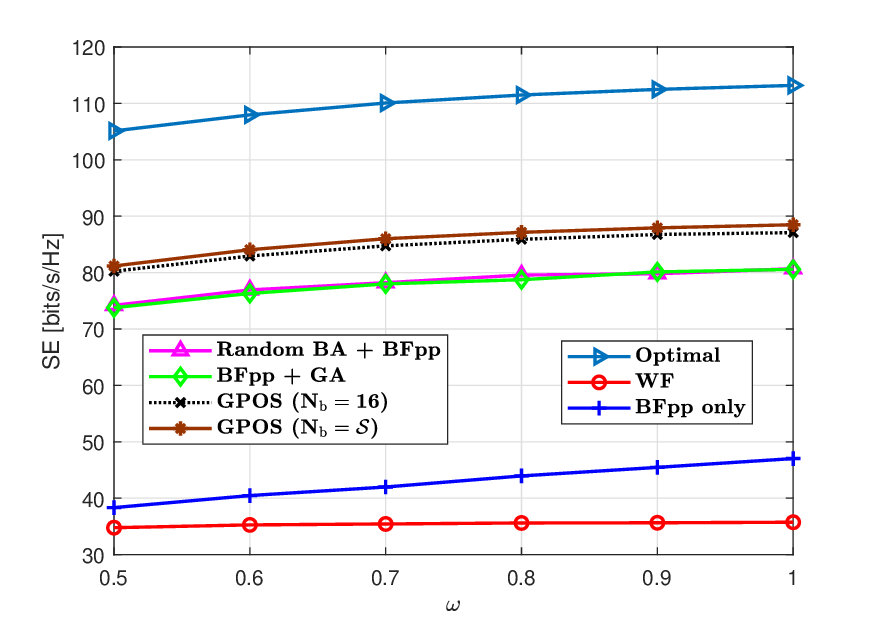}
	\vspace{-2mm}
	\caption{SE performance under imperfect precoder with $\Nt=\Nr=64$, $\Ns=8$, $\textrm{SNR} = 30$~dB, $b=2$, and $\bmax=8$.}
	\label{fig:SE performance versus precoder acc}
\end{figure}

Next, we examine the impact of imperfect precoder on the SE. Specifically, we model the received precoder at the transmitter as $\hat{\Fb}_{\rm t}=\omega \Fb_{\rm r} + \sqrt{1-\omega^2}\Eb$, where $\Fb_{\rm r}$ represents the actual precoder obtained at the receiver, $\omega \in [0,1]$ controls the accuracy of  the precoder feedback, and $\Eb$ models the error with independent $\Ccl\Ncl(0,1)$ entries. The final precoder can be obtained as $\hat{\Fb}_{\rm}=\frac{\sqrt{\Pt}}{\|\hat{\Fb}_{\rm t} \|_{\Fcl}}\hat{\Fb}_{\rm t}$ to satisfy the transmit power budget. Fig.~\ref{fig:SE performance versus precoder acc} shows the SE of the considered schemes under imperfect precoder with $\Nt=\Nr=64$, $\Ns=8$, $\textrm{SNR} = 30$~dB, $b=2$, and $\bmax=8$. We observe that the imperfect precoder affects all the proposed and baseline schemes similarly, resulting in comparable levels of SE degradation. Despite the imperfect precoder, the proposed schemes demonstrate significant SE gains over the baselines, validating their efficiency.

\vspace{-2mm}
\subsection{SE and EE tradeoff}

We herein characterize the SE--EE tradeoff of the considered system. The EE is defined as the ratio between the SE and the total power consumption of the receiver \cite{mendez2016hybrid,abbas2017millimeter}. The latter is given by $P_{\rm total}=  \Nr\left( P_{\rm LNA} + P_{\rm RF}+ 2P_{\rm ADC} \right)$ where $P_{\rm LNA}$, $P_{\rm RF}$, and $P_{\rm ADC}$ denote the power consumption of a low noise amplifier (LNA), an RF chain, and an ADC, respectively. In the following simulations, we set $P_{\rm RF}=43~\mW$ \cite{mendez2016hybrid} and $P_{\rm LNA}=25~\mW$ \cite{gao202022}. Furthermore, a $b$-bit ADC typically has a power consumption of $P_{\rm ADC}=\kappa f_{\rm s}2^b$ \cite{murmann2015race}, where $\kappa$ and $f_{\rm s}$ represent the figure of merit (FoM) and the sampling frequency (ideally equal to the signal bandwidth), respectively. In the simulations, we set the bandwidth to $1$~GHz and choose a conservative value of the FoM, i.e., $\kappa=494$ fJ/step/Hz \cite{abbas2017millimeter}, for evaluation. 
Since beamforming alone with few-bit ADCs is sufficient at low SNRs, as seen from Fig.~\ref{fig:SEvsSNR_b2}, we only consider the high SNR scenarios in the following. Furthermore, the GPOS-BFBA algorithm refers to the ``GPOS ($\Nb=16)$'' scheme.

 \begin{figure}[t]
\small
    \centering
    \hspace{-5mm}
    \subfigure[SE versus $\varsigma$.]
    {\label{fig:SE_varsigmaNr64Nt64Ns8SNR20bit38}\includegraphics[width=0.24\textwidth]{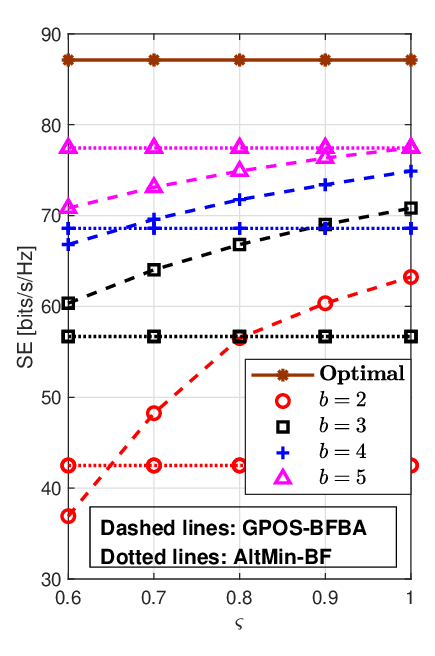}}
     \hspace{-3mm}
    \subfigure[EE versus $\varsigma$.]
    {\label{fig:EE_varsigmaNr4Ns4Nt8SNR20} \includegraphics[width=0.24\textwidth]{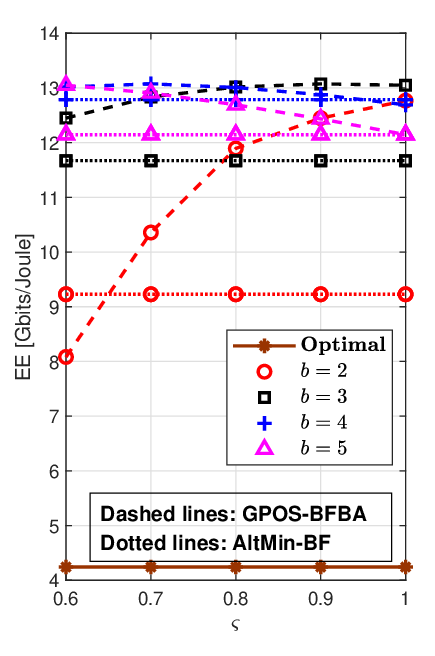}}
    \vspace{-2mm}
    \caption{SE and EE versus $\varsigma$ with $\Nt=\Nr=64$, $\Ns=8$, $\textrm{SNR} = 20$~dB, and $\bmax=5$. }
    \label{fig:Comparision of SE performance versus varsigma evl bmax5}
\end{figure}

In Figs.~\ref{fig:SE_varsigmaNr64Nt64Ns8SNR20bit38} and~\ref{fig:EE_varsigmaNr4Ns4Nt8SNR20}, we plot the SE and EE of the AltMin-BF and GPOS-BFBA designs as functions of $\varsigma$ with $\Nt=\Nr=64$, $\Ns=8$, $\textrm{SNR} = 20$~dB, and $\bmax=5$. We can observe that the joint beamforming and BA design can significantly outperform the beamforming alone in terms of both SE and EE for low-resolution ($2$--$4$ bits) systems. Particularly, for $b=3$, 
the GPOS-BFBA design offers $6\%$ improvements in both the SE and EE, while requiring $40\%$ fewer active ADC bits compared with the AltMin-BF algorithm. Furthermore, when using a total of $128$ bits (i.e., $b=2$ and $\varsigma=1$), the former achieves improvements of $49\%$ in SE and $39\%$ in EE compared to the latter. Additionally, while the full-precision system achieves significantly higher SE than low-resolution ones, the latter attains substantially higher EE. For example, the GPOS-BFBA design for $b=3$ and $\varsigma=1$ achieves $82\%$ of the optimal SE with a $209\%$ improvement in EE compared to the full-precision system.


  \begin{figure*}[tb]
\small
    \centering 
    \hspace{-8mm}
        \subfigure[EE versus ADC resolution.]
    {\label{fig:EE_NrfBitNr64Nt64SNR20bmaxDyn_Apr03} \includegraphics[width=0.45\textwidth]{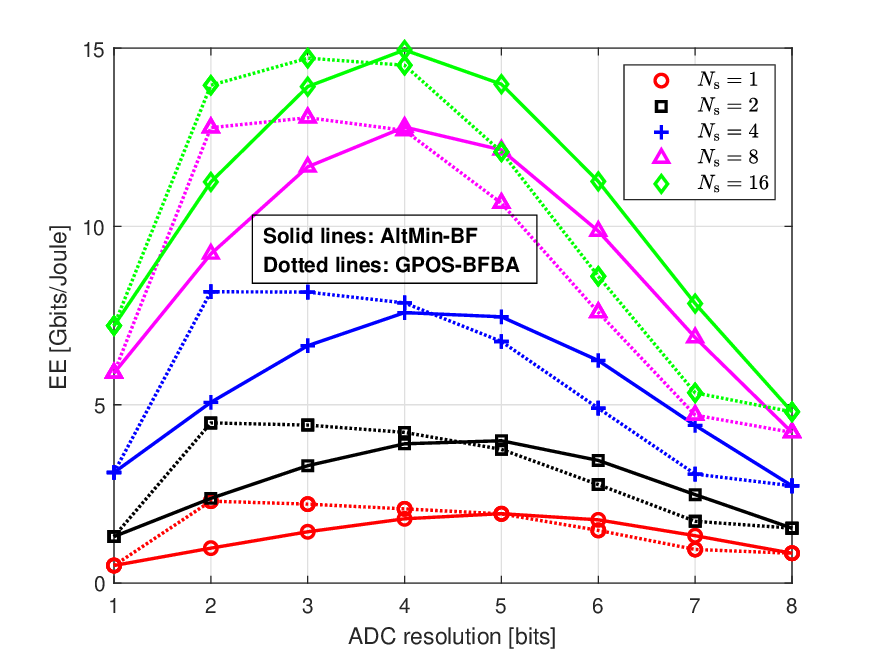}}
            \subfigure[EE-SE map.]
    {\label{fig:EE_SENr64Nt64SNR20bmaxDyn_Apr03} \includegraphics[width=0.45\textwidth]{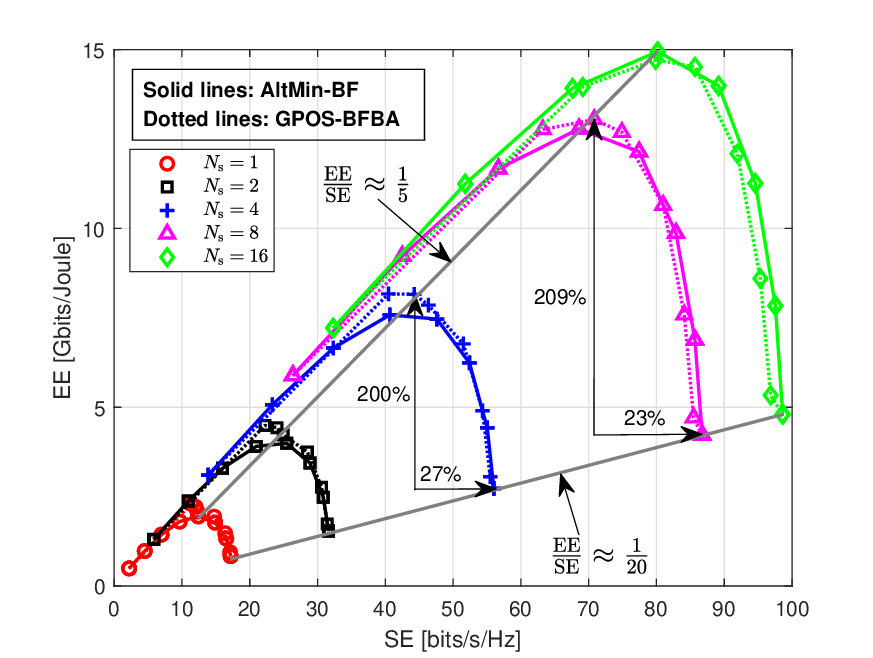}}\hspace{-8mm}
    \vspace{-2mm}
    \caption{SE and EE performance versus ADC resolution ($b$) with $\Nt=\Nr=64$, $\SNR=20~\dB$, and $\varsigma=1$. We set $\bmax=5$ for $b\in\{2,3,4\}$ and $\bmax=b+1$ for $b\in\{5,6,7\}$. The parabolic shape of each line in Fig.~(b) is due to the increase of the ADC bits ranging from 1 to 8.}
    \label{fig:Comparision of SE performance versus Ns evl}
    \vspace{-5mm}
\end{figure*}

Figs.~\ref{fig:EE_NrfBitNr64Nt64SNR20bmaxDyn_Apr03} and \ref{fig:EE_SENr64Nt64SNR20bmaxDyn_Apr03} show the SE and EE of the considered schemes versus the ADC resolution with $\Nt=\Nr=64$, $\SNR=20~\dB$, and $\varsigma=1$. Here, we set $\bmax=5$ for $b\in\{2,3,4\}$ and $\bmax=b+1$ for $b\in\{5,6,7\}$.  It is observed from Fig.~\ref{fig:EE_NrfBitNr64Nt64SNR20bmaxDyn_Apr03} that the AltMin-BF scheme attains both lower SE and EE compared to the GPOS-BFBA design in low-resolution ($2$--$4$ bits) systems. Moreover, we observe that receiving more data streams with low-resolution ADCs can achieve higher SE and EE than receiving fewer data streams with high-resolution ADCs. Table~\ref{tb:An example of SE-EE performance attained by the AltMin-BF scheme.} shows an SE-EE comparison between using $\Ns=4,b=8$ and $\Ns=8,b=3$. Although the SEs are comparable, the EE achieved with the setting $\Ns=8$ and $b=3$ is more than three times that of the setting $\Ns=4$ and $b=8$. Furthermore, the EE-SE map in Fig.~\ref{fig:EE_SENr64Nt64SNR20bmaxDyn_Apr03} shows that more data streams bring forth both higher EE and higher SE. Notably, low-resolution systems that sacrifice less than $30\%$ SE can improve more than $200\%$ EE compared to the full-precision ones. The EE-SE ratios for low-resolution and full-precision systems are approximately $\frac{1}{5}$ and $\frac{1}{20}$, respectively. Therefore, the former can achieve an approximately fourfold improvement in EE compared to the latter for each unit increase in SE.

\begin{table}[hbpt]
\small
\renewcommand\arraystretch{1.2}
\caption{An SE-EE comparison based on the AltMin-BF scheme.}
\centering
    \begin{tabular}{c|c| c} 
    \hline 
      Parameters & SE (bits/s/Hz) &EE (Gbits/Joule)  \\
    \hline  
    $\Ns=4$, $b=8$ &  $56.01$ & $2.73$  \\
    \hline 
    $\Ns=8$, $b=3$ &   $56.09$ & $11.67$\\
        \hline 
    \end{tabular}   
\label{tb:An example of SE-EE performance attained by the AltMin-BF scheme.}
\end{table}

\section{Conclusions}\label{sec:conclusion}

We first establish key properties of optimal quantization, including the scaling law, distortion invariance, and essential statistical characteristics such as the expectation and correlation between a random variable and its quantized output. These properties enable a more accurate characterization of the BAQNM compared to the conventional one \cite{mezghani2012capacity,bai2013optimization} by identifying the underlying conditions, providing a more precise approximation of the distortion factor, and introducing an efficient method for evaluating the QD covariance matrix. The improved modeling accuracy significantly mitigates the performance overestimation commonly observed with conventional BAQNM. Our analytical results reveal that BAQNM and the approximation of the QD covariance matrix typically hold under the condition that the input signal is Gaussian and optimally quantized. Building on these findings, we propose an efficient beamforming design and a low-complexity joint transmit-receive beamforming and bit allocation algorithm that iteratively optimizes both the bit allocation vector and transmit-receive beamforming matrices. Numerical simulations demonstrate the superiority of the proposed schemes over the state-of-the-art designs. Particularly, the proposed joint beamforming and bit allocation design with fewer total ADC bits can achieve both higher SE and EE compared to beamforming alone, especially in low-resolution (2--4 bits) systems. Additionally, the results show that receiving more data streams with low-resolution ADCs can yield higher SE and EE than receiving fewer data streams with high-resolution ADCs. Future work may explore the SE-EE tradeoff under imperfect CSI and beamforming matrices, compare fully digital and hybrid analog-digital architectures, and extend the analysis to wideband multicarrier systems. 
\vspace{-3mm}
\appendices
\section{Proof for Proposition~\ref{lemma:scaling and distortion invariance}}\label{prof:scaling and distortion invariance}
Denote by $\{t_j^{\rm y}, j=0,\ldots,\Nq\} $ and $\{ c_j^{\rm y},j=0,\ldots,\Nq-1\} $ the thresholds and codebook of the optimal quantizer for $Y$, respectively. The MSE between $Y$ and its optimal quantization $\Qy(Y)$ is given by
  \begin{align}\label{eq:quantization MSE Y}
          D_{\rm y}&=\Es\left[\left(\Qy(Y)-Y\right)^2\right]=\sum\limits_{i=0}^{N_{\rm q}-1} \int_{\ty_i}^{\ty_{i+1}} (y-\cy_i)^2 f_Y(y) {\rm d}y\notag \\
          &\overset{(a)}{=}\sigmay^2 \sum\limits_{i=0}^{\Nq-1} \int_{ \frac{\ty_i}{\sigmay} } ^{\frac{\ty_{i+1}}{\sigmay} } (x- \frac{\cy_i}{\sigmay})^2 f_X(x) {\rm d}x,
  \end{align}
  where $(a)$ is due to $f_Y(y)=\frac{1}{\sigma_Y}f_X(\frac{y}{\sigma_Y})$. 
  
  On the other hand, the thresholds and codebook of the optimal quantizer for $X$ are respectively denoted by $\{t_j^{\rm x}, j=0,\ldots,\Nq\} $ and $\{ c_j^{\rm x},j=0,\ldots,\Nq-1\} $. These minimize the quantization MSE of $X$, i.e.,
   \begin{equation}
          D_{\rm x}=\Es\left[\left(\Qx(X)-X\right)^2\right]=\sum\limits_{i=0}^{N_{\rm q}-1} \int_{\tx_i}^{\tx_{i+1}} (x-\cx_i)^2 f_X(x) {\rm d}x.
  \end{equation}

We will show that only when $\cy_i=\sigmay \cx_i$, $D_{\rm y}$ is minimized and satisfies $D_{\rm y}=\sigmay^2 D_{\rm x}$. Specifically, we assume that the codebook of the optimal quantizer for $Y$ satisfies $\cy_i=\sigmay \cx_i + \rho$. Therefore, the optimal threshold satisfies $\ty_i=\sigmay \tx_i + \rho$ according to the nearest neighbor condition. We can rewrite \eqref{eq:quantization MSE Y} as
  \begin{align}\label{eq:Y MSEquan equal}
          D_{\rm y}(\rho)&=\sigmay^2 \sum\limits_{i=0}^{\Nq-1} \int_{ \tx_i+ \frac{\rho}{\sigmay} } ^{\tx_{i+1}+\frac{\rho}{\sigmay} } (x-\cx_i -\frac{\rho}{\sigmay})^2 f_X(x) {\rm d}x, \nonumber\\
          &\overset{(a)}{=}\sigmay^2 \sum\limits_{i=0}^{\Nq-1} \int_{ \tx_i } ^{\tx_{i+1}} (h- \cx_i)^2 f_X(h+\frac{\rho}{\sigmay}) {\rm d}h,
  \end{align}
where $(a)$ is due to $h=x-\frac{\rho}{\sigmay}$. We know that $\{\tx_i,\cx_i\}$ minimizes $D_{\rm y}(0)= \sigmay^2 D_{\rm x}$. As a result, shifting the PDF by $\big| \frac{\rho}{\sigmay} \big|$ changes the contribution to the MSE for each interval $[\tx_i,\tx_{i+1}]$. Since the centroids $\cx_i,\forall i$ are computed to minimize the MSE for the original distribution $f_X(x)$, any shift in the distribution will generally increase the MSE. Therefore, we can conclude that $D_{\rm y}(\rho)\geq D_{\rm y}(0), \forall \rho$, where the equality holds if and only if $\rho=0$.
  
 
\section{Proof for Proposition~\ref{lemma:optimal quantizer condition}}\label{prof:optimal quantizer condition}

For a real zero-mean random variable $Y$, let $Q(Y)$ denote the output of the Lloyd-Max quantizer. The centroid condition \eqref{eq:centriod condition} implies \cite[Ch. 6]{gersho2012vector}:
 \begin{align}
   &\Es[Q(Y)]  =\Es[Y], \label{eq:mean value consistance}\\
   &\Es[Q(Y)(Q(Y)-Y)]  =0.  \label{eq:quantization error uncorrelated to output}
 \end{align}
 Hence, for $X=\Re\{X\}+j \Im\{X\}$,  we have
 \begin{equation}
    \Es[Q(X)]=\Es[Q(\Re\{X\})]+j \Es[Q(\Im\{X\})]=\Es[X].
 \end{equation}
 With $\chi= Q(X)-X$, we can obtain
 \begin{align}
    &  \Es[Q(X)\chi^*]=\Es[Q(X)(Q(X)-X)^*] = 0,
 \end{align}
 where the last equality follows \eqref{eq:quantization error uncorrelated to output} and the assumption that $\Re\{X\}$ and $\Im\{X\} $ are i.i.d and independently quantized. Because $\Re\{X\}$ and $\Im\{X\} $ have the same variance of $\sigma_X^2/2$, we have
 \begin{align}
     \Es[|X|^2 & =2\Es[\Re\{X\}^2]= 2\Es[\Im\{X\}^2],\\
     \Es[|\chi|^2] & =2\Es[\Re\{\chi\}^2]=2\Es[\Im\{\chi\}^2],
 \end{align}
 which yields
 \begin{equation}
 \gamma=\frac{\Es[\left|\chi \right|^2]}{\Es[\left|X\right|^2]}=\frac{\Es[\Re\{\chi\}^2]}{\Es[\Re\{X\}^2]}=\frac{\Es[\Im\{\chi\}^2]}{\Es[\Im\{X\}^2]}
 \end{equation}
 in Proposition~\ref{lemma:optimal quantizer condition}.

 \vspace{-2mm}
 \section{Proof for Corollary~\ref{lemma:approximated Cov of eta and z}}\label{prof:approximated Cov of eta and z}
  \vspace{-2mm}
 The quantization error $q_m=z_m-y_m$, conditioned on $y_m$, is statistically independent of all other random variables of the system. Hence, for $m\neq n$, we have
 \begin{align}\label{eq:qmqn correlation}
\Es[q_m q_n^*] & =\Es\left[\Es[q_m q_n^*|y_n] \right]=\Es\left[ \Es[q_m |y_n]  \Es[q_n^* |y_n]   \right]  \nonumber \\
& \overset{(a)}{\approx} \! \Es \! \left[ C_{q_m,y_n}C_{y_n}^{-1}y_n \Es \! [q_n^* |y_n]   \right] \! = \! C_{q_m,y_n}C_{y_n}^{-1} \Es \! \left[ y_n q_n^*  \right],
 \end{align}
 where $(a)$ is due to the LMMSE estimation.
  Furthermore, we have
 \begin{align}\label{eq:cross-correlation of qm xn}
     C_{q_m,y_n}&=\Es\left[q_m y_n^*\right]=\Es\left[(z_m-y_m) y_n^*\right]=C_{z_m,y_n}-C_{y_m,y_n} \nonumber \\
     &\overset{(c)}{=}(g_m-1) C_{y_m,y_n} \overset{(d)}{=}-\gamma_m C_{y_m,y_n},
 \end{align}
 where (c) and (d) are due to Lemma~\ref{lemma:element-wise quantization} and Corollary~\ref{eq:Bussgang gain}, respectively. Similarly, we obtain
 \begin{align}\label{eq: cross-correlation of xn qn}
     \Es\left[ y_n q_n^*  \right]&=\Es\left[ y_n (z_n-y_n)^*\right]=C_{z_n,y_n}^*-C_{y_n} \nonumber \\
     &=C_{y_n} (g_n^*-1)  =-\gamma_n C_{y_n}.
 \end{align}
  Based on \eqref{eq:qmqn correlation}--\eqref{eq: cross-correlation of xn qn}, we obtain $\Es[q_m q_n^*]\approx \gamma_m \gamma_n \Es[y_m y_n^*]$. Combined with $\Es[q_n q_n^*]=\gamma_n\Es[y_n y_n^*]$, we have
 \begin{align}
     \Cb_{q}&\approx\diag(\Cb_{y})\Gammab + \Gammab \, \nondiag(\Cb_{y})\Gammab \nonumber \\
     &=\Gammab\Cb_{y}\Gammab+(\Ib-\Gammab)\diag(\Cb_{y})\Gammab,
 \end{align}
 where $\nondiag(\Ab)$ denotes the matrix containing all non-diagonal entries of $\Ab$ while its diagonal entries are all zero. Based on \eqref{eq:CoV of error term with q} and \eqref{eq:Cov of z with q}, we have
 \begin{align}
 & \Cb_{\eta} \approx \Gammab \, \diag(\Cb_{y})(\Ib-\Gammab), \\
 & \Cb_{z} \approx \left[\diag(\Cb_{y})\Gammab +(\Ib-\Gammab)\Cb_{y}\right](\Ib-\Gammab).
 \end{align}

It is observed from the results $\Es[q_m q_n^*]\approx \gamma_m \gamma_n \Es[y_m y_n^*]$ and $\Es[q_n q_n^*]=\gamma_n\Es[y_n y_n^*]$ that the cross-correlation coefficient is obtained by the LMMSE estimation while the auto-correlation coefficient comes from the definition of the distortion factor. Therefore, the diagonal entries of $\Cb_{\eta}$ are exactly the ones of $\Gammab \, \diag(\Cb_{y})(\Ib-\Gammab)$. As such, the approximation is due to neglecting the non-zero off-diagonal entries of $\Cb_{\eta}$.

\section{Proof for Proposition~\ref{prop:WMMSE equivalence DBF}}\label{sec:appendices}

With the first-order condition of local optima, we obtain
  \vspace{1mm}
\begin{align}
	& \Wb =\Eb^{-1}, \label{eq:solution to W initial} \\
	& \Ub =\left(\Gb\Hb\Fb\Fb^\H \Hb^\H \Gb + \Cb_e \right)^{-1}\Gb\Hb \Fb. \label{eq:solution to U proof}
\end{align}
Therefore, the MSE matrix can be recast as
\begin{equation}
	\Eb=\Ib-\Fb^\H \Hb^\H \Gb \left(\Gb\Hb\Fb\Fb^\H \Hb^\H \Gb + \Cb_e \right)^{-1} \Gb \Hb \Fb.
\end{equation}
Using the Woodbury matrix identity, we obtain
\begin{equation}\label{eq:solution to W refined proof}
	\Eb^{-1}=\Ib + \Fb^\H \Hb^\H\Gb \Cb_e^{-1} \Gb \Hb \Fb=\Wb.
\end{equation}
Therefore, we have
\begin{align}
	\hspace{-1mm} f(\Ub, \Fb,\Wb) & =\Nt-\log \det(\Eb^{-1}) \nonumber \\
	& =\Nt-\log \det\left(\Ib + \Cb_e^{-1} \Gb \Hb \Fb\Fb^\H \Hb^\H\Gb \right).
\end{align}
We next show that $R=\det\left(\Ib + \Cb_e^{-1} \Gb \Hb \Fb\Fb^\H \Hb^\H\Gb \right)$ with $\Ub$ given by the MMSE solution \eqref{eq:solution to U proof}. With $\Lb=\Gb \Hb \Fb$, $\Ub$ can be written as $\Ub=\left(\Lb\Lb^\H+ \Cb_e\right)^{-1}\Lb$ based on \eqref{eq:solution to U proof}. Furthermore, by the Woodbury matrix identity, we obtain
\begin{equation}\label{eq:U equivalent form}
	\Ub=\Cb_e^{-1}\Lb -\Cb_e^{-1}\Lb\left(\Ib +\Lb^\H \Cb_e^{-1}\Lb\right)^{-1}\Lb^\H \Cb_e^{-1}\Lb,
\end{equation}
which results in
\begin{subequations}\label{eq:Ub product}
	\begin{align}
		& \Ub^\H\Cb_e =\left(\Ib - \Pb (\Ib +\Pb) \right)^{-1}\Lb^\H\Ub=(\Ib+\Pb)^{-1}\Lb^\H\Ub, \\
		& \Lb^\H\Ub =\Pb\left( \Ib -\Pb(\Ib+\Pb)^{-1}\right)=\Pb(\Ib+\Pb)^{-1},
	\end{align}
\end{subequations}
where $\Pb= \Lb^\H \Cb_e^{-1}\Lb$ and we note that $(\Ib+\Pb)^{-1}=\Ib-(\Ib+\Pb)^{-1}\Pb=\Ib-\Pb(\Ib+\Pb)^{-1}$. With \eqref{eq:Ub product}, we derive
\begin{align}
	R&=\log \det \left( \Ib +  (\Ub^\H\Cb_e\Ub)^{-1}\Ub^\H \Lb \Lb^\H \Ub\right)=\log \det \left( \Ib +  \Pb\right) \nonumber \\
	&=\log \det\left(\Ib + \Cb_e^{-1} \Gb \Hb \Fb\Fb^\H \Hb^\H\Gb \right) \nonumber \\
 &=\Nt-f(\Ub, \Fb,\Wb).
\end{align}
Hence, problem \eqref{pb:precoder design WMMSE} is equivalent to \eqref{pb:subproblem of Fb and Ub} when $\Ub$ and $\Wb$ are given by \eqref{eq:solution to U} and \eqref{eq:solution to W refined}, respectively. 

\bibliographystyle{IEEEtran}
\bibliography{conf_short,jour_short,refs-my}

\begin{thebibliography}{10}
\providecommand{\url}[1]{#1}
\csname url@samestyle\endcsname
\providecommand{\newblock}{\relax}
\providecommand{\bibinfo}[2]{#2}
\providecommand{\BIBentrySTDinterwordspacing}{\spaceskip=0pt\relax}
\providecommand{\BIBentryALTinterwordstretchfactor}{4}
\providecommand{\BIBentryALTinterwordspacing}{\spaceskip=\fontdimen2\font plus
\BIBentryALTinterwordstretchfactor\fontdimen3\font minus
  \fontdimen4\font\relax}
\providecommand{\BIBforeignlanguage}[2]{{%
\expandafter\ifx\csname l@#1\endcsname\relax
\typeout{** WARNING: IEEEtran.bst: No hyphenation pattern has been}%
\typeout{** loaded for the language `#1'. Using the pattern for}%
\typeout{** the default language instead.}%
\else
\language=\csname l@#1\endcsname
\fi
#2}}
\providecommand{\BIBdecl}{\relax}
\BIBdecl

\bibitem{heath2016overview}
R.~W. Heath, N.~Gonzalez-Prelcic, S.~Rangan, W.~Roh, and A.~M. Sayeed, ``An
  overview of signal processing techniques for millimeter wave {MIMO}
  systems,'' \emph{{IEEE} J. Sel. Areas Commun.}, vol.~10, no.~3, pp. 436--453,
  2016.

\bibitem{jiang2021road}
W.~Jiang, B.~Han, M.~A. Habibi, and H.~D. Schotten, ``The road towards {6G}: A
  comprehensive survey,'' \emph{{IEEE} Open J. Commun. Soc.}, vol.~2, pp.
  334--366, 2021.

\bibitem{li2017channel}
Y.~Li, C.~Tao, G.~Seco-Granados, A.~Mezghani, A.~L. Swindlehurst, and L.~Liu,
  ``Channel estimation and performance analysis of one-bit massive {MIMO}
  systems,'' \emph{{IEEE} Trans. Signal Process.}, vol.~65, no.~15, pp.
  4075--4089, 2017.

\bibitem{murmann2015race}
B.~Murmann, ``The race for the extra decibel: A brief review of current {ADC}
  performance trajectories,'' \emph{{IEEE} Solid-State Circuits Mag.}, vol.~7,
  no.~3, pp. 58--66, 2015.

\bibitem{liu2019low}
J.~Liu, Z.~Luo, and X.~Xiong, ``Low-resolution {ADCs} for wireless
  communication: A comprehensive survey,'' \emph{{IEEE} Access}, vol.~7, pp.
  91\,291--91\,324, 2019.

\bibitem{mendez2016hybrid}
R.~M{\'e}ndez-Rial, C.~Rusu, N.~Gonz{\'a}lez-Prelcic, A.~Alkhateeb, and R.~W.
  Heath, ``Hybrid {MIMO} architectures for millimeter wave communications:
  Phase shifters or switches?'' \emph{{IEEE} Access}, vol.~4, pp. 247--267,
  2016.

\bibitem{roth2018comparison}
K.~Roth, H.~Pirzadeh, A.~L. Swindlehurst, and J.~A. Nossek, ``A comparison of
  hybrid beamforming and digital beamforming with low-resolution {ADCs} for
  multiple users and imperfect {CSI},'' \emph{{IEEE} J. Sel. Topics Signal
  Process.}, vol.~12, no.~3, pp. 484--498, 2018.

\bibitem{yan2019performance}
H.~Yan, S.~Ramesh, T.~Gallagher, C.~Ling, and D.~Cabric, ``Performance, power,
  and area design trade-offs in millimeter-wave transmitter beamforming
  architectures,'' \emph{{IEEE} Circuits Syst. Mag.}, vol.~19, no.~2, pp.
  33--58, 2019.

\bibitem{castaneda2021resolution}
O.~Casta{\~n}eda, Z.~Boynton, S.~H. Mirfarshbafan, S.~Huang, C.~Y. Jamie,
  A.~Molnar, and C.~Studer, ``A resolution-adaptive 8 mm$^2$ 9.98 {Gb/s} 39.7
  {pJ/b} 32-antenna all-digital spatial equalizer for {mmWave} massive
  {MU-MIMO} in 65nm {CMOS},'' in \emph{Proc. Solid-State Circ. Conf.}, 2021.

\bibitem{tse2005fundamentals}
D.~Tse and P.~Viswanath, \emph{Fundamentals of wireless communication}.\hskip
  1em plus 0.5em minus 0.4em\relax Cambridge university press, 2005.

\bibitem{singh2009limits}
J.~Singh, O.~Dabeer, and U.~Madhow, ``On the limits of communication with
  low-precision analog-to-digital conversion at the receiver,'' \emph{{IEEE}
  Trans. Commun.}, vol.~57, no.~12, pp. 3629--3639, 2009.

\bibitem{jacobsson2017throughput}
S.~Jacobsson, G.~Durisi, M.~Coldrey, U.~Gustavsson, and C.~Studer, ``Throughput
  analysis of massive {MIMO} uplink with low-resolution {ADCs},'' \emph{{IEEE}
  Trans. Wireless Commun.}, vol.~16, no.~6, pp. 4038--4051, 2017.

\bibitem{mezghani2012capacity}
A.~Mezghani and J.~A. Nossek, ``Capacity lower bound of {MIMO} channels with
  output quantization and correlated noise,'' in \emph{Proc. IEEE Int. Symp.
  Inf. Theory}, 2012.

\bibitem{mezghani2009transmit}
A.~Mezghani, R.~Ghiat, and J.~A. Nossek, ``Transmit processing with low
  resolution {D/A}-converters,'' in \emph{Proc. IEEE Int. Conf. Electron.,
  Circuits, Syst.}, 2009.

\bibitem{jacobsson2017quantized}
S.~Jacobsson, G.~Durisi, M.~Coldrey, T.~Goldstein, and C.~Studer, ``Quantized
  precoding for massive {MU-MIMO},'' \emph{{IEEE} Trans. Wireless Commun.},
  vol.~65, no.~11, pp. 4670--4684, 2017.

\bibitem{ling2019performance}
X.~Ling and R.~Wang, ``Performance analysis and transceiver design of few-bit
  quantized {MIMO} systems,'' \emph{{IEEE} Access}, vol.~7, pp. 9935--9944,
  2019.

\bibitem{ma2024digital}
M.~Ma, N.~T. Nguyen, I.~Atzeni, A.~L. Swindlehurst, and M.~Juntti, ``Digital
  and hybrid precoding designs in massive {MIMO} with low-resolution {ADCs},''
  \emph{arXiv preprint arXiv:2409.17638}, 2024.

\bibitem{zhang2016mixed}
T.-C. Zhang, C.-K. Wen, S.~Jin, and T.~Jiang, ``{Mixed-ADC} massive {MIMO}
  detectors: Performance analysis and design optimization,'' \emph{{IEEE}
  Trans. Wireless Commun.}, vol.~15, no.~11, pp. 7738--7752, 2016.

\bibitem{zhang2017performance}
J.~Zhang, L.~Dai, Z.~He, S.~Jin, and X.~Li, ``Performance analysis of
  mixed-{ADC} massive {MIMO} systems over {Rician} fading channels,''
  \emph{{IEEE} J. Sel. Areas Commun.}, vol.~35, no.~6, pp. 1327--1338, 2017.

\bibitem{pirzadeh2018spectral}
H.~Pirzadeh and A.~L. Swindlehurst, ``Spectral efficiency of {mixed-ADC}
  massive {MIMO},'' \emph{{IEEE} Trans. Signal Process.}, vol.~66, no.~13, pp.
  3599--3613, 2018.

\bibitem{bai2013optimization}
Q.~Bai, A.~Mezghani, and J.~A. Nossek, ``On the optimization of {ADC}
  resolution in multi-antenna systems,'' in \emph{Proc. Int. Symp. Wireless
  Commun. Systems}, 2013.

\bibitem{ahmed2017joint}
I.~Z. Ahmed, H.~Sadjadpour, and S.~Yousefi, ``A joint combiner and bit
  allocation design for massive {MIMO} using genetic algorithm,'' in
  \emph{Proc. Asilomar Conf. Signals, Syst., Comp.}, 2017.

\bibitem{choi2017resolution}
J.~Choi, B.~L. Evans, and A.~Gatherer, ``Resolution-adaptive hybrid {MIMO}
  architectures for millimeter wave communications,'' \emph{{IEEE} Trans.
  Signal Process.}, vol.~65, no.~23, pp. 6201--6216, 2017.

\bibitem{nguyen2020energy}
K.-G. Nguyen, Q.-D. Vu, L.-N. Tran, and M.~Juntti, ``Energy-efficient bit
  allocation for resolution-adaptive {ADC} in multiuser large-scale {MIMO}
  systems: Global optimality,'' in \emph{Proc. IEEE Int. Conf. Acoust., Speech,
  and Signal Process.}, 2020.

\bibitem{prasad2020optimizing}
N.~Prasad, X.~F. Qi, and A.~Molev-Shteiman, ``Optimizing resolution-adaptive
  massive {MIMO} networks,'' in \emph{Proc. IEEE Int. Conf. on Comp. Commun.},
  2020.

\bibitem{castaneda2021spawc}
O.~Casta{\~n}eda, S.~H. Mirfarshbafan, S.~Ghajari, A.~Molnar, S.~Jacobsson,
  G.~Durisi, and C.~Studer, ``Resolution-adaptive all-digital spatial
  equalization for {mmWave} massive {MU-MIMO},'' in \emph{Proc. IEEE Workshop
  Signal Proc. Adv. in Wirel. Comms.}, 2021.

\bibitem{sheng2020energy}
H.~Sheng, X.~Chen, X.~Zhai, A.~Liu, and M.-J. Zhao, ``Energy efficiency
  optimization for millimeter wave system with resolution-adaptive {ADCs},''
  \emph{{IEEE} Wireless Commun. Lett.}, vol.~9, no.~9, pp. 1519--1523, 2020.

\bibitem{verenzuela2021optimal}
D.~Verenzuela, E.~Bj{\"o}rnson, and M.~Matthaiou, ``Optimal per-antenna {ADC}
  bit allocation in correlated and cell-free massive {MIMO},'' \emph{{IEEE}
  Trans. Commun.}, vol.~69, no.~7, pp. 4767--4780, 2021.

\bibitem{verenzuela2017per}
------, ``Per-antenna hardware optimization and mixed resolution {ADCs} in
  uplink massive {MIMO},'' in \emph{Proc. Asilomar Conf. Signals, Syst.,
  Comp.}, 2017.

\bibitem{gersho2012vector}
A.~Gersho and R.~M. Gray, \emph{Vector quantization and signal
  compression}.\hskip 1em plus 0.5em minus 0.4em\relax Springer Science \&
  Business Media, 2012, vol. 159.

\bibitem{fletcher2007robust}
A.~K. Fletcher, S.~Rangan, V.~K. Goyal, and K.~Ramchandran, ``Robust predictive
  quantization: Analysis and design via convex optimization,'' \emph{{IEEE} J.
  Sel. Topics Signal Process.}, vol.~1, no.~4, pp. 618--632, 2007.

\bibitem{orhan2015low}
O.~Orhan, E.~Erkip, and S.~Rangan, ``Low power analog-to-digital conversion in
  millimeter wave systems: Impact of resolution and bandwidth on performance,''
  in \emph{Proc. ITG Workshop Smart Antennas}, 2015.

\bibitem{bussgang1952crosscorrelation}
J.~J. {Bussgang}, ``Crosscorrelation functions of amplitude-distorted
  {Gaussian} signals,'' Res. Lab. Electron., Massachusetts Inst. Technol.,
  Tech. Rep. 216, 1952.

\bibitem{diggavi2001worst}
S.~N. Diggavi and T.~M. Cover, ``The worst additive noise under a covariance
  constraint,'' \emph{{IEEE} Trans. Inf. Theory}, vol.~47, no.~7, pp.
  3072--3081, 2001.

\bibitem{hassibi2003much}
B.~Hassibi and B.~M. Hochwald, ``How much training is needed in
  multiple-antenna wireless links?'' \emph{{IEEE} Trans. Inf. Theory}, vol.~49,
  no.~4, pp. 951--963, 2003.

\bibitem{demir2020bussgang}
O.~T. Demir and E.~Bjornson, ``The {Bussgang} decomposition of nonlinear
  systems: Basic theory and {MIMO} extensions [lecture notes],'' \emph{{IEEE}
  Signal Process. Mag.}, vol.~38, no.~1, pp. 131--136, 2020.

\bibitem{roth2017achievable}
K.~Roth and J.~A. Nossek, ``Achievable rate and energy efficiency of hybrid and
  digital beamforming receivers with low resolution {ADC},'' \emph{{IEEE} J.
  Sel. Areas Commun.}, vol.~35, no.~9, pp. 2056--2068, 2017.

\bibitem{jacovitti1994estimation}
G.~Jacovitti and A.~Neri, ``Estimation of the autocorrelation function of
  complex gaussian stationary processes by amplitude clipped signals,''
  \emph{{IEEE} Trans. Inf. Theory}, vol.~40, no.~1, pp. 239--245, 1994.

\bibitem{lin2019transceiver}
Y.~Lin, S.~Jin, M.~Matthaiou, and X.~You, ``Transceiver design with {UCD}-based
  hybrid beamforming for millimeter wave massive {MIMO},'' \emph{{IEEE} Trans.
  Commun.}, vol.~67, no.~6, pp. 4047--4061, 2019.

\bibitem{ning2023beamforming}
B.~Ning, Z.~Tian, W.~Mei, Z.~Chen, C.~Han, S.~Li, J.~Yuan, and R.~Zhang,
  ``Beamforming technologies for ultra-massive {MIMO} in terahertz
  communications,'' \emph{{IEEE} Open J. Commun. Soc.}, vol.~4, pp. 614--658,
  2023.

\bibitem{dai2022delay}
L.~Dai, J.~Tan, Z.~Chen, and H.~V. Poor, ``Delay-phase precoding for wideband
  {THz} massive {MIMO},'' \emph{{IEEE} Trans. Wireless Commun.}, vol.~21,
  no.~9, pp. 7271--7286, 2022.

\bibitem{qi2022hybrid}
C.~Qi, Q.~Liu, X.~Yu, and G.~Y. Li, ``Hybrid precoding for mixture use of phase
  shifters and switches in {mmWave} massive {MIMO},'' \emph{{IEEE} Trans.
  Commun.}, vol.~70, no.~6, pp. 4121--4133, 2022.

\bibitem{zhang2017hybridly}
D.~Zhang, Y.~Wang, X.~Li, and W.~Xiang, ``Hybridly connected structure for
  hybrid beamforming in {mmWave} massive {MIMO} systems,'' \emph{{IEEE} Trans.
  Commun.}, vol.~66, no.~2, pp. 662--674, 2017.

\bibitem{gao2021wideband}
F.~Gao, B.~Wang, C.~Xing, J.~An, and G.~Y. Li, ``Wideband beamforming for
  hybrid massive {MIMO} terahertz communications,'' \emph{{IEEE} J. Sel. Areas
  Commun.}, vol.~39, no.~6, pp. 1725--1740, 2021.

\bibitem{yoo2002power}
J.~Yoo, D.~Lee, K.~Choi, and J.~Kim, ``A power and resolution adaptive flash
  analog-to-digital converter,'' in \emph{Proc. ACM Int. Symp. Low Power
  Electron. Des.}, 2002.

\bibitem{nahata2004high}
S.~Nahata, K.~Choi, and J.~Yoo, ``A high-speed power and resolution adaptive
  flash analog-to-digital converter,'' in \emph{Proc. IEEE Int. Syst.-Chip
  Conf.}, 2004.

\bibitem{rajashekar2008design}
G.~Rajashekar and M.~Bhat, ``Design of resolution adaptive tiq flash adc using
  ams 0.35 $\mu$m technology,'' in \emph{Proc. IEEE Int. Conf. Electron. Des},
  2008.

\bibitem{mo2015capacity}
J.~Mo and R.~W. Heath, ``Capacity analysis of one-bit quantized {MIMO} systems
  with transmitter channel state information,'' \emph{{IEEE} Trans. Signal
  Process.}, vol.~63, no.~20, pp. 5498--5512, 2015.

\bibitem{wang2022channel}
Y.~Wang, X.~Chen, Y.~Cai, B.~Champagne, and L.~Hanzo, ``Channel estimation for
  hybrid massive {MIMO} systems with adaptive-resolution {ADCs},'' \emph{{IEEE}
  Trans. Commun.}, vol.~70, no.~3, pp. 2131--2146, 2022.

\bibitem{rangan2014millimeter}
S.~Rangan, T.~S. Rappaport, and E.~Erkip, ``Millimeter-wave cellular wireless
  networks: Potentials and challenges,'' \emph{Proc. {IEEE}}, vol. 102, no.~3,
  pp. 366--385, 2014.

\bibitem{yu2016alternating}
X.~Yu, J.-C. Shen, J.~Zhang, and K.~B. Letaief, ``Alternating minimization
  algorithms for hybrid precoding in millimeter wave {MIMO} systems,''
  \emph{{IEEE} J. Sel. Topics Signal Process.}, vol.~10, no.~3, pp. 485--500,
  2016.

\bibitem{bjornson2018hardware}
E.~Bj{\"o}rnson, L.~Sanguinetti, and J.~Hoydis, ``Hardware distortion
  correlation has negligible impact on {UL} massive {MIMO} spectral
  efficiency,'' \emph{{IEEE} Trans. Commun.}, vol.~67, no.~2, pp. 1085--1098,
  2018.

\bibitem{max1960quantizing}
J.~Max, ``Quantizing for minimum distortion,'' \emph{{IRE} Trans. Inf. Theory},
  vol.~6, no.~1, pp. 7--12, 1960.

\bibitem{usman2016mmse}
O.~B. Usman, H.~Jedda, A.~Mezghani, and J.~A. Nossek, ``{MMSE} precoder for
  massive {MIMO} using 1-bit quantization,'' in \emph{Proc. IEEE Int. Conf.
  Acoust., Speech, and Signal Process.}, 2016.

\bibitem{atzeni2021channel}
I.~Atzeni and A.~T{\"o}lli, ``Channel estimation and data detection analysis of
  massive {MIMO} with 1-bit {ADCs},'' \emph{{IEEE} Trans. Wireless Commun.},
  vol.~21, no.~6, pp. 3850--3867, 2021.

\bibitem{berger2010application}
C.~R. Berger, Z.~Wang, J.~Huang, and S.~Zhou, ``Application of compressive
  sensing to sparse channel estimation,'' \emph{{IEEE} Commun. Mag.}, vol.~48,
  no.~11, pp. 164--174, 2010.

\bibitem{lee2016channel}
J.~Lee, G.-T. Gil, and Y.~H. Lee, ``Channel estimation via orthogonal matching
  pursuit for hybrid {MIMO} systems in millimeter wave communications,''
  \emph{{IEEE} Trans. Commun.}, vol.~64, no.~6, pp. 2370--2386, 2016.

\bibitem{chae2008TSP}
C.-B. Chae, D.~Mazzarese, T.~Inoue, and R.~W. Heath, ``Coordinated beamforming
  for the multiuser {MIMO} broadcast channel with limited feedforward,''
  \emph{{IEEE} Trans. Signal Process.}, vol.~56, no.~12, pp. 6044--6056, 2008.

\bibitem{el2014spatially}
O.~El~Ayach, S.~Rajagopal, S.~Abu-Surra, Z.~Pi, and R.~W. Heath, ``Spatially
  sparse precoding in millimeter wave {MIMO} systems,'' \emph{{IEEE} Trans.
  Wireless Commun.}, vol.~13, no.~3, pp. 1499--1513, 2014.

\bibitem{akdeniz2014millimeter}
M.~R. Akdeniz, Y.~Liu, M.~K. Samimi, S.~Sun, S.~Rangan, T.~S. Rappaport, and
  E.~Erkip, ``Millimeter wave channel modeling and cellular capacity
  evaluation,'' \emph{{IEEE} J. Sel. Areas Commun.}, vol.~32, no.~6, pp.
  1164--1179, 2014.

\bibitem{jacobsson2019linear}
S.~Jacobsson, G.~Durisi, M.~Coldrey, and C.~Studer, ``Linear precoding with
  low-resolution {DACs} for massive {MU-MIMO-OFDM} downlink,'' \emph{IEEE
  Transactions on Wireless Communications}, vol.~18, no.~3, pp. 1595--1609,
  2019.

\bibitem{chu2019efficient}
L.~Chu, F.~Wen, L.~Li, and R.~Qiu, ``Efficient nonlinear precoding for massive
  {MIMO} downlink systems with 1-bit {DACs},'' \emph{{IEEE} Trans. Wireless
  Commun.}, vol.~18, no.~9, pp. 4213--4224, 2019.

\bibitem{abbas2017millimeter}
W.~B. Abbas, F.~Gomez-Cuba, and M.~Zorzi, ``Millimeter wave receiver
  efficiency: A comprehensive comparison of beamforming schemes with low
  resolution {ADCs},'' \emph{{IEEE} Trans. Wireless Commun.}, vol.~16, no.~12,
  pp. 8131--8146, 2017.

\bibitem{gao202022}
L.~Gao and G.~M. Rebeiz, ``A {22--44-GHz} phased-array receive beamformer in
  45-nm {CMOS SOI} for {5G} applications with 3--3.6-db {NF},'' \emph{{IEEE}
  Trans. Microw. Theory Techn.}, vol.~68, no.~11, pp. 4765--4774, 2020.

\end{thebibliography}

\end{document}